\documentclass[a4paper,fleqn,usenatbib,useAMS]{mnras}

\usepackage{graphicx}	
\usepackage{amsmath}	
\usepackage{amssymb}	
\usepackage{multicol}        
\usepackage{bm}		
\usepackage{pdflscape}	
\usepackage{times}
\usepackage{mathtools}
\usepackage{float}
\usepackage{xcolor}

\newcommand{\HII}{H\,{\sc ii}}
\newcommand{\NII}{[N\,{\sc ii}]}
\newcommand{\OIII}{[O\,{\sc iii}]}
\newcommand{\OII}{[O\,{\sc ii}]}

\newcommand{\Ha}{H$\alpha$}
\newcommand{\Hb}{H$\beta$}

\newcommand{\Mo}{~M$_{\odot}$}
\newcommand{\Msun}{~M$_{\odot}$}

\newcommand{\Mstar}{${\rm M}_{\rm star}$}
\newcommand{\Mbar}{M$_{\rm bar}$}

\newcommand{\Mb}{${\rm M}_{\rm bar}$}

\usepackage[T1]{fontenc}
\usepackage{ae,aecompl}

\usepackage{newtxtext,newtxmath}

\title[Oxygen yields as feedback indicators]{Oxygen yields as a constraint on feedback processes in galaxies}

\author[Maritza A. Lara-L\'opez]{Maritza A. Lara-L\'opez$^{1}$\thanks{E-mail: maritza@dark-cosmology.dk}, Maria Emilia De Rossi$^{2,3}$, Leonid S. Pilyugin$^{4}$, Anna Gallazzi$^{5}$,
\newauthor
Thomas M. Hughes$^{6,7,8,9}$, Igor A. Zinchenko$^{4, 10}$\\
$^{1}$DARK, Niels Bohr Institute, University of Copenhagen, Lyngbyvej 2, Copenhagen DK-2100, Denmark\\
$^{2}$Universidad de Buenos Aires, Facultad de Ciencias Exactas y Naturales y Ciclo Básico Común. Buenos Aires, Argentina\\
$^{3}$CONICET-Universidad de Buenos Aires, Instituto de Astronomía y Física del Espacio (IAFE). Buenos Aires, Argentina\\
 $^{4}$Main Astronomical Observatory, National Academy of Sciences of Ukraine, 27 Akademika Zabolotnoho St, 03680 Kiev, Ukraine\\
$^{5}$INAF -- Osservatorio Astrofisico di Arcetri, Largo Enrico Fermi 5, I-50125 Firenze, Italy\\
$^{6}$ Chinese Academy of Sciences South America Center for Astronomy, China-Chile Joint Center for Astronomy, Camino El Observatorio \#1515, Las Condes, Santiago, Chile\\
$^{7}$ Instituto de F\'{i}sica y Astronom\'{i}a, Universidad de Valpara\'{i}so, Avda. Gran Breta\~{n}a 1111, Valpara\'{i}so, Chile\\
$^{8}$ CAS Key Laboratory for Research in Galaxies and Cosmology, Department of Astronomy, University of Science and Technology of
China, Hefei 230026, China\\
$^{9}$ School of Astronomy and Space Science, University of Science and Technology of China, Hefei 230026, China\\
$^{10}$Astronomisches Rechen-Institut, Zentrum für Astronomie der Universität Heidelberg, Mönchhofstr. 12-14, 69120, Heidelberg, Germany\\}

\date{Accepted XXX. Received YYY; in original form ZZZ}

\pubyear{2019}

\begin{document}
\label{firstpage}
\pagerange{\pageref{firstpage}--\pageref{lastpage}}
\maketitle

\begin{abstract}

We study the interplay between several properties determined from Optical and a combination of Optical/Radio measurements, such as  the effective Oxygen yield (y$_{\rm eff}$), the star formation efficiency, gas metallicity, depletion time, gas fraction, and baryonic mass (\Mbar), among others. We use  spectroscopic data from the SDSS survey,  and HI information from the ALFALFA survey to build a statistically significant sample of more than 5,000 galaxies. Furthermore, we complement our analysis with data from the GASS and COLD GASS surveys, and with a sample of star forming galaxies from the Virgo cluster.  Additionally, we have compared our results with predictions from the EAGLE simulations, finding a very good agreement when using the high resolution run. We explore in detail the \Mbar-y$_{\rm eff}$ relation, finding a  bimodal trend that can be separated when the stellar age of galaxies is considered. On one hand, y$_{\rm eff}$  increases with \Mbar\ for young galaxies(log(t$_{\rm r}$) $<$ 9.2 yr),  while y$_{\rm eff}$ shows an anti-correlation with \Mbar\ for older galaxies (log(t$_{\rm r}$) $>$ 9.4 yr). While a correlation between  \Mbar\ and y$_{\rm eff}$ has been observed and studied before, mainly for samples of dwarfs and irregular galaxies, their anti-correlated counterpart for massive galaxies has not been previously reported. The EAGLE simulations indicate that AGN feedback must have played an important role in their history by quenching their star formation rate, whereas low mass galaxies would have been affected by a combination of outflows and infall of gas.

\end{abstract}




\section{Introduction}

The evolution of galaxies is intimately dependent on the conversion of gas into stars, the production of heavy elements, recycling of this material into the interstellar medium, and repetitions of this cycle. A detailed understanding of the interplay between each of gas mass, star formation rate, and metallicity is clearly important to understand the galaxy evolution process. 

A key variable in the gas recycling process and a primary element in the evolution of galaxies is the amount of atomic and molecular hydrogen, or HI and H$_2$, respectively. Large HI surveys such as the Arecibo Legacy Fast Arecibo L-band Feed Array \citep[ALFALFA, ][]{Haynes11}, and the GALEX Arecibo SDSS Survey \citep[GASS, ][]{Catinella10}, have provided HI information for thousands and hundreds of galaxies, respectively, leading to new scaling relationships and dependencies in galaxies. Additionally, CO information using the IRAM 30-m telescope has been explored by the COLD GASS survey \citep{Saintonge11}, obtaining H$_2$ estimates for a sample of $\sim$350 nearby, massive galaxies.

Scaling relations between stellar mass and star formation rate (M-SFR), and between mass and metallicity (M-Z), have been explored for many years  \citep[e.g., ][]{Tremonti04, Noeske07, Lara13a}. Only during the last decade and with the advent of large surveys, multiwavelength approaches combining radio and optical observations have successfully provided critical information to investigate the connection between neutral and/or molecular gas and metallicity for large samples of galaxies. For example, \citet{Both13} analyze the relationship between metallicity, gas content and stellar mass of local galaxies using data from SDSS and ALFALFA, finding a fundamental relationship between those variables. \citet{Lara13b} studied the relationship between cold gas, metallicity and specific SFR through the Z-SSFR relationship. They found that at a given mass, and depending on the amount of gas, galaxies follow opposite behaviors; while low-mass galaxies with high/low gas fractions show high/moderate SSFR and low/high metallicities, respectively, whereas high-mass galaxies with moderate/low gas fractions will show high/low metallicities.

Different physical explanations can be given to justify the nature of scaling relationships. Many observational and theoretical approaches have been done around the origin of the M-Z relation, either attributed to the presence of inflows \citep{Dalcan07}, outflows \citep{Larson74}, differences in the star formation efficiency (SFE) of galaxies \citep{Brooks09, Fin08, Calura09}, or variations on the initial mass function \citep[IMF,][]{Koppen07}.

However, the degree in which each one of them affect the metal enrichment and SFR in galaxies is still a matter of debate. For instance, \citet[][]{Calura09}  reproduced the M-Z relation with chemical evolution models by increasing the efficiency of star formation in galaxies of all morphological types, without the need for outflows favoring the loss of metals in the less massive galaxies. Furthermore, \citet[][]{Vale09} modeled the time evolution of stellar metallicity using a closed-box chemical evolution picture. They suggest that the M-Z relation for galaxies in the mass range from 10$^{9.8}$ to 10$^{11.65 }$\Mo\  is mainly driven by the star formation history and not by inflows or outflows.

On the other hand, it was pointed out in the high-resolution simulations of \citet[][]{Brooks09}, that supernovae feedback plays a crucial role in lowering the star formation efficiency in low-mass galaxies. Without energy injection from supernovae to regulate the star formation, gas that remains in galaxies rapidly cools, forms stars, and increases its metallicity too early, producing a M-Z relation that is too flat compared to observations. Nevertheless, \citet[][]{Dalcan07} affirm that the fraction of baryonic mass lost through winds is quite modest ($<$15\%) at all galaxy masses. Supernova feedback is therefore unlikely to be effective for removing large amounts of gas from low-mass disk galaxies.

Another important feedback process that likely took place in the history of current massive SF galaxies is given by AGNs. By using EAGLE simulations \citet[][]{derossi2017} showed that the slope of the M-Z relation is mainly modulated by stellar feedback at low masses, whilst  AGN feedback regulates the slope at the high-mass end. In this scenario, AGN feedback would generate a decrease in the global metallicity of SF gas by quenching the star formation activity of simulated galaxies, and also by ejecting metal-enriched material.

This paper will make use of effective yields, which may indicate observationally whether a galaxy has experienced inflows and/or outflows. A galaxy that evolves as a closed box \citep[e.g.,][]{PP75} obeys a simple analytic relationship between the metallicity of the gas (Z$_{\rm gas}$), and the gas mass fraction $\mu$= M$_{\rm gas}$ / (M$_{\rm gas}$ + M$_{\rm star})$, where M$_{\rm gas}$ is the gas mass, and  M$_{\rm star}$ is the stellar mass (hereafter, M$_{\rm star}$ implies M$_{\rm star}/$M$_{\rm sun}$). As gas is converted into stars, the gas mass fraction decreases and the metallicity of the gas increases according to \citet[][]{SS72} as:
\begin{equation}\label{EqYield}
{\rm Z}_{\rm gas}={y}_{\rm true} {\rm ln}(1/\mu)
\end{equation} where $y_{\rm true}$ is the true nucleosynthetic yield, defined as the mass in primary elements freshly produced by massive stars, in units of stellar mass that is locked up in long-lived stars and stellar remnants. If a galaxy evolves as a closed box, the ratio of Eq.  \ref{EqYield} should be a constant equal to the nucleosynthetic yield. However, this ratio will be lower if metals have been lost from the system through outflows of gas, or if the current gas has been diluted with fresh infall of metal-poor gas. Therefore, the next ratio has been defined as effective yield:
\begin{equation}\label{EqEffYield}
y_{\rm eff}={\rm Z}_{\rm gas}/{\rm In} (1/\mu)
\end{equation}

The effective yield will be constant ($y_{\rm eff}$=$y_{\rm true}$) for any galaxy that has evolved as a closed box. The simple closed-box model assumes both instantaneous recycling and that the products of stellar nucleosynthesis are neither diluted by infalling pristine gas nor lost to the system via outflow of enriched gas \citep[e.g.,][]{Talbot71, SS72}.
 In contrast, if any gas has either entered or left the galaxy, the effective yield will drop below the closed box value due to changes in metallicity and/or gas mass fraction. For example, by means of an analytical chemical evolution model, \citep[e.g.,][]{Kudritzki15} showed that  high effective yield of galaxies may be explained by relatively low rates of accretion and winds.  Therefore, the effective yield is an observational quantity that can be used to diagnose the addition or removal of gas in a galaxy, see \citet[][]{Edmunds90} and \citet[][]{Dalcan07}  for a more detailed explanation on yields.

\citet[][]{Tremonti04} used a sample of $\sim$ 53,000 star forming galaxies from the SDSS at z $\sim$ 0.1. By using indirect estimates of the gas mass based on the Ha luminosity, they found that y$_{\rm eff}$ decreases with decreasing baryonic mass, attributing the decrease of the effective yield at low galaxy masses to galactic winds removing metals more efficiently from the shallower potential wells of dwarf galaxies.

The direct measurement  of oxygen yields requires radio observations to measure gas masses, as well as spectroscopy to measure metallicities. Then, it is natural  that so far direct estimations of oxygen yields have been explored in relatively small samples. For example, by investigating the maximum metallicity value of spiral galaxies, \citet[][]{Pil07} determined the gas metallicity through the electron temperature, obtaining an oxygen yield of about $\sim$ 0.0035. By comparing observations with analytical models,  \citet[][]{Dalcan07} reached 3 main conclusions: (1) metal-enriched outflows are the only mechanism that can significantly reduce the effective yield, but only for gas rich systems. (2) It is nearly impossible to reduce the effective yield of a gas-poor system, no matter how much gas is lost or accreted, and (3) any subsequent star formation drives the effective yield back to the close box value. 


Several authors have explored effective oxygen yields for samples of dwarfs and irregular galaxies. For instance, using a sample of dwarf and irregular galaxies,  \citet[][]{Ekta10} analyzed extremely metal poor (XMP) galaxies , finding that the effective oxygen yield increases with increasing baryonic mass, consistent with what is expected if outflows of metal-enriched gas are important in determining the effective yield. Also, galaxies in their sample deviate from the luminosity-metallicity relation due to a combination of being gas rich, and having a more uniform mixing of metals in their ISM. On the same line, \citet[][]{SA14,SA15}, have argued that the low metallicities of the XMD galaxies are an indicator of infall of pristine gas. This process would increase the effective yield of low-mass galaxies as compared to more massive galaxies.

Furthermore, \citet[][]{Garnett02} analyzed a sample of 40 spiral and irregular galaxies with rotation speeds ranging from a few to 300 km/s. He finds that the effective yield correlates with V$_{\rm rot}$ up to V$_{\rm rot}$ $\sim$ 125 km/s. However, the V$_{\rm rot}$-Y$_{\rm eff}$ relation shows a change in slope for  higher V$_{\rm rot}$  values, where there appears to be no relation.  \citet[][]{Garnett02} attributes this change in behavior to increasing loss of metals from the smaller galaxies in supernova-driven winds.

\citet[][]{Thuan16} analyzed a sample of 29 extremely metal-deficient blue compact dwarf (BCD) galaxies, finding that about 65$\%$ of the BCDs in their sample have an y$_{\rm eff}$ larger than the y$_{\rm true}$. They explain that the most likely scenario is that the  y$_{\rm eff}$ has been overestimated, since when calculating it, it is assumed that the metallicity of the neutral gas is equal to that of the ionized gas, an assumption not likely to be true. They show that the neutral gas envelope in BCDs is more metal-deficient by a factor   of $\sim$ 1.5-20, as compared to the ionized gas \citep[see also][]{Filho13, Thuan05}. The rest 35$\%$ of galaxies with effective yields lower than the true yield, can be understood as the result of the loss of metals due to supernova-driven outflows, and/or the consequence dilution by inflows of metal-poor gas. They find, however, that in their sample there is not an evident variation of the effective yield with the baryonic mass of the galaxy, as found by \citet[][]{Garnett02, Tremonti04, Ekta10}


Additionally,  \citet[][]{Hughes13} analyzed the link between the gas metallicity and HI content in galaxies. They investigated the role of cold gas and environment using 260 nearby late--type galaxies from isolated to Virgo cluster members, finding a constant average effective yield of $y_{\rm eff}=10^{-2.6}$ with a 0.2 dex scatter across the mass range. In general, they observe that gas--poor galaxies are typically more metal rich, and demonstrate that the removal of gas from the outskirts of spirals increases the observed average metallicity by $\sim$ 0.1 dex. 





On the other hand, spatially resolved analysis of oxygen yields have emerge in the last years. For instance, \citet{Vilchez19} find that the metal budget of NGC 628 and M 101 disk appears consistent with the predictions of the simple model of chemical evolution for an oxygen yield between half and one solar, although M 101 present deviations in the outermost region, suggesting the presence of gas flow \citep[see also][for an analysis of M31]{Telford19}. Also, \citet[][]{Zasov15} analyzed the radial distribution of the effective yield for a sample of 14 spiral galaxies. They show that the maximal values of the effective yield in the main disks of galaxies anticorrelate with the total mass of galaxies and with the mass of their dark halos enclosed within R$_{25}$.


In this paper, we examine several scaling relations by combining several optical and radio properties, in particular for oxygen yields. This paper is organized as follows: In $\S\,\ref{SampleSelection}$ we detail the data used for this study, in   $\S\,$ 3 we analyze scaling relations by combining properties derived from optical and radio data. In $\S\,$ 4, we analyze and compare our results with those obtained with the EAGLE simulations. In  $\S\,$ 5 the analysis of the relationship between \Mbar\ and ${\rm y}_{\rm eff}$ is presented. A discussion is presented in $\S\,$ 6, and finally, our conclusions are given in $\S\,\ref{Conclusion}$. Throughout this paper we assume $H_0=70\,$km\,s$^{-1}$\,Mpc$^{-1}$, $\Omega_M=0.3$, $\Omega_{\Lambda}=0.7$.

\section[]{Sample selection}\label{SampleSelection}

The combination of panchromatic data represent a major step on the understanding of how galaxies assemble their mass and evolve. This paper combines optical and radio wavelengths to produce combined properties, such as gas fractions ($\mu$), SFE, baryonic masses (M$_{\rm bar}$), and effective oxygen yields (${\rm y}_{\rm eff}$).  

We consider data for emission-line galaxies from several large surveys. In Optical, we use the  Sloan Digital Sky Survey--Data Release 7 \citep[SDSS--DR7,][]{Abaza09} and the spectroscopic sample of the Virgo cluster from \citep{Hughes13}. In radio we use the ALFALFA  \citep{Haynes11} and GASS \citep{Catinella10} surveys to obtain HI gas, and the COLD GASS \citep{Saintonge11} survey to obtain H$_2$ information for a subsample of our galaxies. All galaxies from GASS and COLD GASS surveys have optical counterparts from the SDSS survey.   Since our sample is taken from different surveys, it present different spatial resolutions and instrument aperture sizes. A more detailed analysis  will be given using EAGLE simulations in Appendix \ref{Ap2}.


\subsection[]{Derived properties from optical data}\label{OpticalData}

Data from the SDSS were taken with the 2.5 m telescope located at Apache Point Observatory \citep{Gunn06}. We use the emission-line analysis of SDSS-DR7 galaxy spectra performed by the MPA-JHU  database\footnote{http://www.mpa-garching.mpg.de/SDSS/}. From the full dataset, we only consider objects classified as galaxies in the $``$main galaxy sample$"$ \citep{Strauss02} with apparent Petrosian $r$--magnitudes in the range $14.5 < m_{r} < 17.77$. Additionally, we make use of the spectroscopic data from the Virgo cluster taken with the CARELEC spectrograph on the 1.93 m telescope at Observatoire de Haute Provence, as described by \citet{Hughes13}.

For reliable metallicity, SFRs, and E(B-V) estimates, we selected galaxies with a signal-to-noise ratio (SNR) of 3 in  {H$\alpha$}, {H$\beta$},  [{N\,\textsc{ii}}], and  [{O\,\textsc{iii}}] $\lambda$5007 (Fig. \ref{BPT_HI}, grey sample). Additionally, we selected only SF galaxies using the standard  BPT diagram \citep{Baldwin81}, with the discrimination of \citet{Kauf03b}. Our final SF optical sample for SDSS is of 252,513 galaxies (Fig. \ref{BPT_HI}, blue sample). This sample is used to match with ALFALFA, GASS, and COLDGASS data as described in $\S\,$  \ref{Radiodata}.

\begin{figure}\label{BPT_HI}
\begin{center}
\includegraphics[scale=0.5]{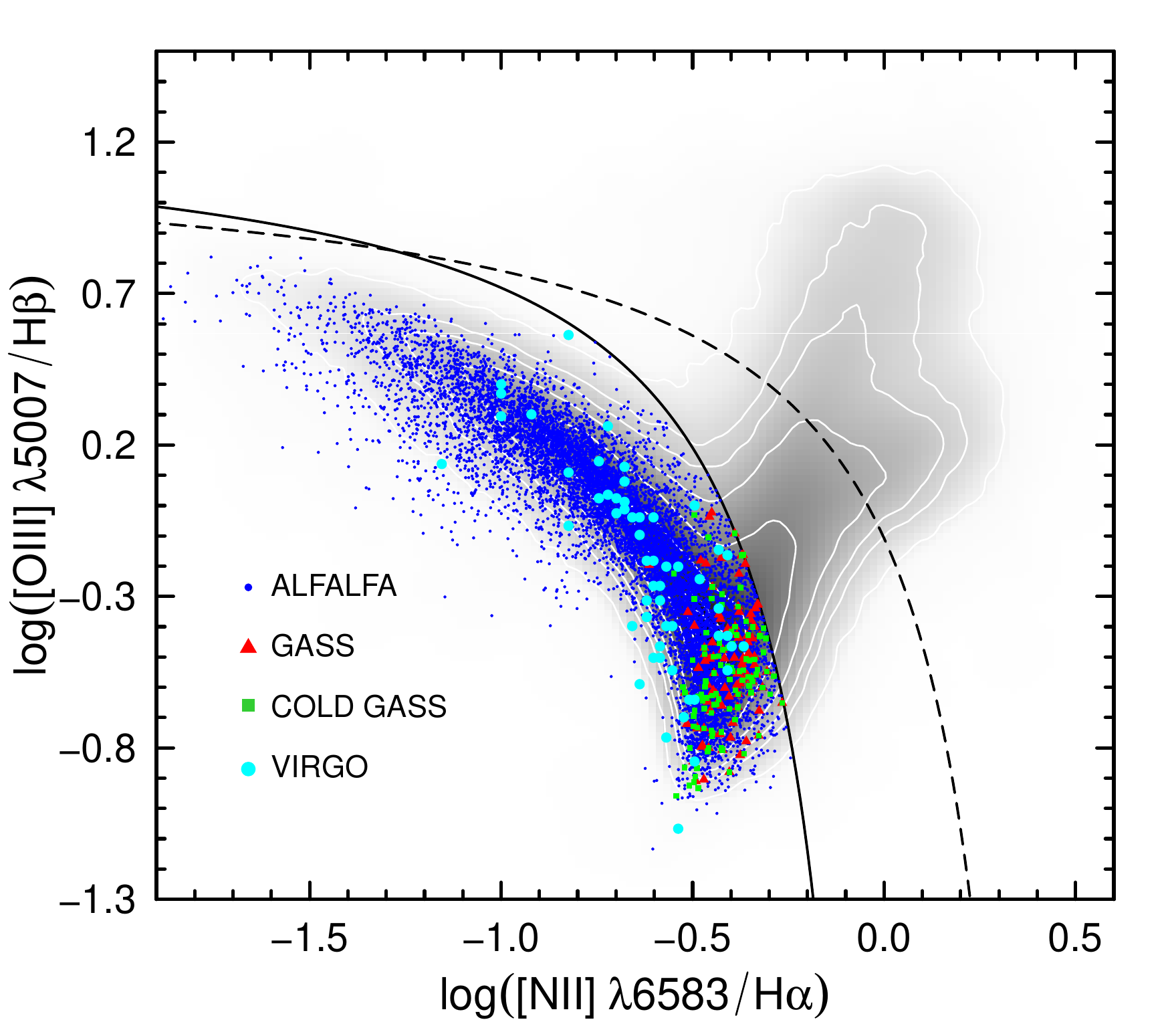}
\caption{BPT diagram for our final sample of galaxies. Grey background data show SDSS galaxies with a S/N $>$ 3 in the four emission lines plotted. SDSS star forming galaxies with counterparts in the AFALFA survey are shown in blue, with the GASS survey are shown in red triangles, and COLD GASS survey in green squares.  Galaxies from the Virgo cluster are shown in cyan filled circles.}
\label{BPT_HI}
\end{center}
\end{figure}

For SDSS galaxies, total stellar masses were taken from the MPA-JHU database, and are described in \citet{Kauf03a}. Also, SFRs and  Specific Star Formation Rates (SSFRs) were taken from the same database, and are described by \citet{Brinchmann04}.   In order to identify possible biases,  we compare \citet{Brinchmann04}  SFRs with two different estimations:  \citet{Duarte17}, and direct integrated SFRs from the MaNGA-DR14. We find a very good agreement between them, with a mean difference between MaNGA and \citet{Brinchmann04} of $\sim$0.026,  and between MaNGA and \citet{Duarte17} of  $\sim$0.03. For further discussion, refer to  Appendix \ref{Ap1}.\\
 For Virgo cluster galaxies, we  estimated SFRs based on the {H$\alpha$} line following the prescription of  \citet{Hopkins03} and assuming a Salpeter IMF. SFRs are corrected by Balmer Decrements, stellar absorption, aperture and obscuration as detailed by \citet{Brough11}, \citet{Mad11}, and \citet{Lara13a}. SFRs were converted to a Chabrier et al. IMF dividing them by 1.5, and  converted to the \citet{Brinchmann04} system applying the  linear conversion from \citet{Lara13a}:

\begin{equation}\label{ConvSFR}
{\rm log(SFR)}_{\rm BCH}=0.09845 + 0.90806\times \rm{log(SFR)}_{\rm HOP}
\end{equation}

SFRs for Virgo cluster galaxies are also estimated using the FUV data as described by \citet{Hughes13}. Since SFRs estimations using {H$\alpha$} and FUV agree very well, to keep consistency with the data used in this paper, we use SFRs estimated with the {H$\alpha$} line.

To estimate gas metallicities, we use the three-dimensional calibration relations for abundance determinations from a set of strong emission lines suggested by  \citet{Pil16}. The oxygen abundances 12+log(O/H)$_{\rm R}$ are determined using the R calibration (Eq. \ref{RcalUp}), which uses the following ratios: R$_2$ = {\OII} ${\rm \lambda 3727 + \lambda 3729}$ / ${\rm{H\beta}}$, R$_3$ =  {\OIII} ${\rm{\lambda 4959 + \lambda 5007}}$ / ${\rm{H\beta}}$, and  N$_2$ =  {\NII} ${\rm \lambda 6548 + \lambda 6584}$ / ${\rm{H\beta}}$.  \\
The R calibration is bimodal, for log(N$_2$) $\geq$ $-$0.6 (the upper branch), we have:
\begin{equation}\label{RcalUp}
       \begin{split}
& {\rm 12+log(O/H)_{R,U}} = 8.589 + 0.022\; {\rm log(R_3 / R_2)} + 0.399\; {\rm log (N_2)}  \\
& + \;[\; - 0.137 + 0.164\; {\rm log(R_3 / R_2)} + 0.589 \; {\rm log(N_2)}\;] \times {\rm log (R_2)} \\ \\
& {\rm while\; for\; log(N_2) < -0.6 \; (lower\; branch):} \; \\
& {\rm 12+log(O/H)_{R,L}} = 7.932 + 0.944\; {\rm log(R_3 / R_2)} + 0.695\; {\rm log (N_2)}  \\
& + \;[\; 0.970 + 0.291\; {\rm log(R_3 / R_2)} - 0.019 \; {\rm log(N_2)}\;] \times {\rm log (R_2)} \\
       \end{split}
\end{equation}

It has been argued that this three-dimensional R calibration produces reliable abundances for fibre spectra from the SDSS  \citep{Pil18}. It should be noted that the oxygen abundances determined through the R calibration are compatible to the metallicity scale of {\HII}  regions defined by the  {\HII} regions with abundances obtained through the direct $T_e$ method.  Note that for SDSS data we are estimating gas metallicities in a single 3" fiber spectra per galaxy. To address the biases due to the SDSS fiber, in Appendix \ref{Ap1} we compare SDSS single fiber and integrated MaNGA IFU data for 59 common sources, finding a mean difference  of $\sim$0.03 dex in 12+log(O/H).

 Since the Virgo sample (and a sub-sample of SDSS galaxies) do not have all the emission lines to estimate directly the R calibration, we created a calibration using 7733 SDSS objects with the oxygen abundances determined through the R calibration of \citet{Pil18}, we derived the following calibration for the O3N2 index:

\begin{equation} \label{Rrecal}
\rm {12+log(O/H) }_{\rm R_{O3N2}} = 8.546 (\pm 0.001) - 0.265  (\pm 0.001) \times \rm O3N2
\end{equation} where the O3N2 index is defined as \citep{Pettini04}:  


\begin{equation} \label{O3N2}
\rm O3N2 =  log\left({\frac{[{\rm O\,\textsc{iii}}] \;  \lambda 5007/{\rm H}\beta}{[{\rm N\,\textsc{ii}}]  \; \lambda 6583/{\rm H}\alpha}}\right),
\end{equation}

 The mean difference between the metallicities estimated directly using \citet{Pil16} method, and metallicities estimated from   Eq. \ref{Rrecal} is   $\sim$ 0.01 dex, while the mean absolute percentage error (100$\% / n$ $\times$ $\sum_{i=1}^{n} | \ (y_i-\hat{y_i})/y_i$ | , where $y$ is the actual value and $\hat{y}$ the predicted) is  $\sim$ 0.62$\%$.

The oxygen abundance estimated from the emission lines reflects the gas-phase oxygen abundance. Nevertheless, some fraction of the oxygen is locked into dust grains. The dust depletion of oxygen increases with metallicity and is around 0.12 dex for {\HII} regions of the Orion metallicity  \citep[12 + log(O/H) $\sim$ 8.5, ][]{Mesa09, PP10,Espiritu17}. We adopt the following simplest expression to describe the dust depletion of oxygen as a function of the gas-phase abundance:

\begin{equation} \label{Metd}
\rm {log(O/H) }_{\rm D} = 0.08 \times [ 12+log(O/H)_R - 7.0 ]
\end{equation} 

Then, the total gas + dust oxygen abundance, used throughout this paper, is:

\begin{equation} \label{MetT}
\rm {12+log(O/H) }_{\rm T} = 12+log(O/H)_R + log(O/H)_D
\end{equation} 

Finally, as an indicator of dust content, we estimated the color excess E(B-V) \citep[e.g.][]{Calzetti96}. For its estimation, different radios from Hydrogen recombination lines can be used, for our sample, we are using \Ha \ and \Hb, and the prescription of \citet{Calzetti94} as follows:

\begin{equation}
{\rm E\,(B-V)}_{\rm gas} =  {\frac{ {\rm log} ({\rm R}_{\rm obs} / {\rm R}_{\rm int}) \;  }{ 0.4 \, [ \, k (\lambda_{{\rm H}\alpha}) -  k  (\lambda_{{\rm H}\beta})\,] }}
\end{equation}

where the observed radio ${\rm R}_{\rm obs}$ is the Balmer decrement  (\Ha/\Hb), and the intrinsic ratio ${\rm R}_{\rm int}$ is 2.86, taking a case B recombination  \citep[][]{Osterbrock89}. Finally, $k(\lambda)$ corresponds to the dust extinction curve, in this case, we use \citet{Calzetti2000}, which is based on a set of starburst galaxies, appropriate for this analysis:

\begin{equation}
       \begin{split}
& {k}({\lambda}) =  2.659 \; \left(\; - 1.857 +  {\frac {1.040 } { \lambda}  }\right) + 4.05,\\
& {\rm for} \;  0.63\;  \mu m \leq \lambda \leq 2.2\;  \mu m, \; {\rm and}\\
& {k}({\lambda}) =  2.659 \; \left(\; - 2.156 +  {\frac {1.509 } { \lambda} } - {\frac {0.198 } { \lambda^2} } + {\frac {0.011 } { \lambda^3} } \right) + 4.05,\\
& {\rm for} \;  0.12\;  \mu m \leq \lambda \leq 0.63\;  \mu m\\
       \end{split}
\end{equation}

%


\subsection[]{Combined properties from optical and radio data}\label{Radiodata}

ALFALFA \citep[][]{Haynes11} is a blind survey of 21 cm HI emission  over $\sim$ 2800 deg$^2$ of sky. We are using the catalog public version\footnote{http://egg.astro.cornell.edu/alfalfa/data/index.php} $\alpha$.70, which provides $\sim$ 15,855 HI detections, of which $\sim$ 15,041 are associated with extragalactic objects. Since ALFALFA is a blind survey, we select only galaxies with Code=1, which refers to robust, reliably detected sources. We also remove sources with heliocentric velocities V$_{helio}$$<$100.0, which are unlikely to be galaxies. From this subsample, we cross-match RA, DEC, and redshift for our optical sample, obtaining a total of 15,049 galaxies, from which 11,610 are classified as SF galaxies and meet the SN required described in $\S\,$\ref{OpticalData}.

In order to increase our HI coverage of massive galaxies, we match our SDSS SF--sample described in  $\S\,$\ref{OpticalData} with the GASS public catalogue. The GASS survey observed a sample of $\sim$ 666 galaxies selected from the SDSS spectroscopic and GALEX imaging surveys. GASS galaxies are selected to have stellar masses greater than 10$^{10}$ M$_{\rm star}$ and redshifts 0.025$<z<$0.05. We use the public catalogue of HI detections, which includes 379 galaxies, from which a total  of  102 galaxies belong to our  SF--sample described in $\S\,$ \ref{OpticalData}. The rest of the GASS galaxies correspond to composite and AGN galaxies. 

Galaxies from the COLD GASS survey \citep{Saintonge11} were included as well in our sample. We found a total of 366 matches with SDSS data, from which 102 are star forming galaxies, as shown in Fig. 1. Additionally, we included galaxies from Virgo with HI detections from \citep{Hughes13}, adding a total of 55 SF galaxies to our sample.

For SDSS/ALFALFA, SDSS/GASS, and Virgo galaxies,  H$_2$ gas masses (M$_{\rm H2}$) were calculated using the prescription of \citep{Saintonge11} as follows:

\begin{equation}\label{EqYield22}
{\rm log}({\rm M_{H2}}/ \rm M_{HI})={0.45 \, [log(\rm M_{\rm star})-10.7]\, -0.387} 
\end{equation} 

For galaxies with information from the COLD GASS survey, we used the M$_{\rm H2}$ estimated directly through CO as shown in \citet{Saintonge11}.\\
For the scaling relations analyzed in this paper, gas masses are estimated as M$_{\rm gas}$ $=$ M$_{\rm HI}$ + M$_{\rm H2}$,  baryonic masses as M$_{\rm bar}$ $=$ M$_{\rm gas}$ + M$_{\rm star}$, and gas fractions as:

\begin{equation}\label{GasFracEq}
{\rm \mu }= {\rm M}_{\rm gas}/{\rm (M}_{\rm gas}+{\rm M}_{\rm star})
\end{equation}

Effective yields were estimated using Eq. \ref{EqEffYield} as ${\rm y}_{\rm eff}={\rm Z}_{\rm gas}/{\rm In}(1/\mu)$. The value of the oxygen abundance O/H is traditionally expressed in units of the number of oxygen atoms relative to hydrogen, while the value of Z$_{\rm gas}$ in Eq. \ref{EqEffYield} have units of mass fraction, we adopt the following relation given by \citet{Garnett02}:

\begin{equation}\label{EqZ}
{\rm Z_{\rm gas}} = 12 \times {\frac {O } { H }   }
\end{equation} where O/H is obtained from Eq. \ref{MetT}

Finally, we proceed to estimate the SFE. There are many definitions for the SFE, taking the ratio of the SFR with respect to either the molecular gas,  the neutral gas, or the gas mass  \citep[e.g.][]{Leroy08}. Throughout this paper we are using the global SFE, defined as  SFE $=$ SFR/M$_{\rm  gas}$. Additionally, we are estimating the global depletion time  as  t$_{\rm dep}$ $=$ M$_{\rm gas}$ / SFR. A previous study analyzing the different depletion times for M$_{\rm H2}$ and M$_{\rm HI}$ was performed by \citet{Saintonge11}.

For all the estimated variables in $\S\,$ \ref{OpticalData} and  \ref{Radiodata}, we estimated errors using  the ``Propagate" package developed for the ``R" statistical programming language. This package perform propagation of uncertainty using higher-order Taylor expansion and Monte Carlo simulation. We consistently used the errors provided by the Monte Carlo simulations for all our variables.






\subsection[]{Stellar mass completeness}\label{MassComplet}

\begin{figure}\label{CompSample}
\begin{center}
\includegraphics[scale=0.5]{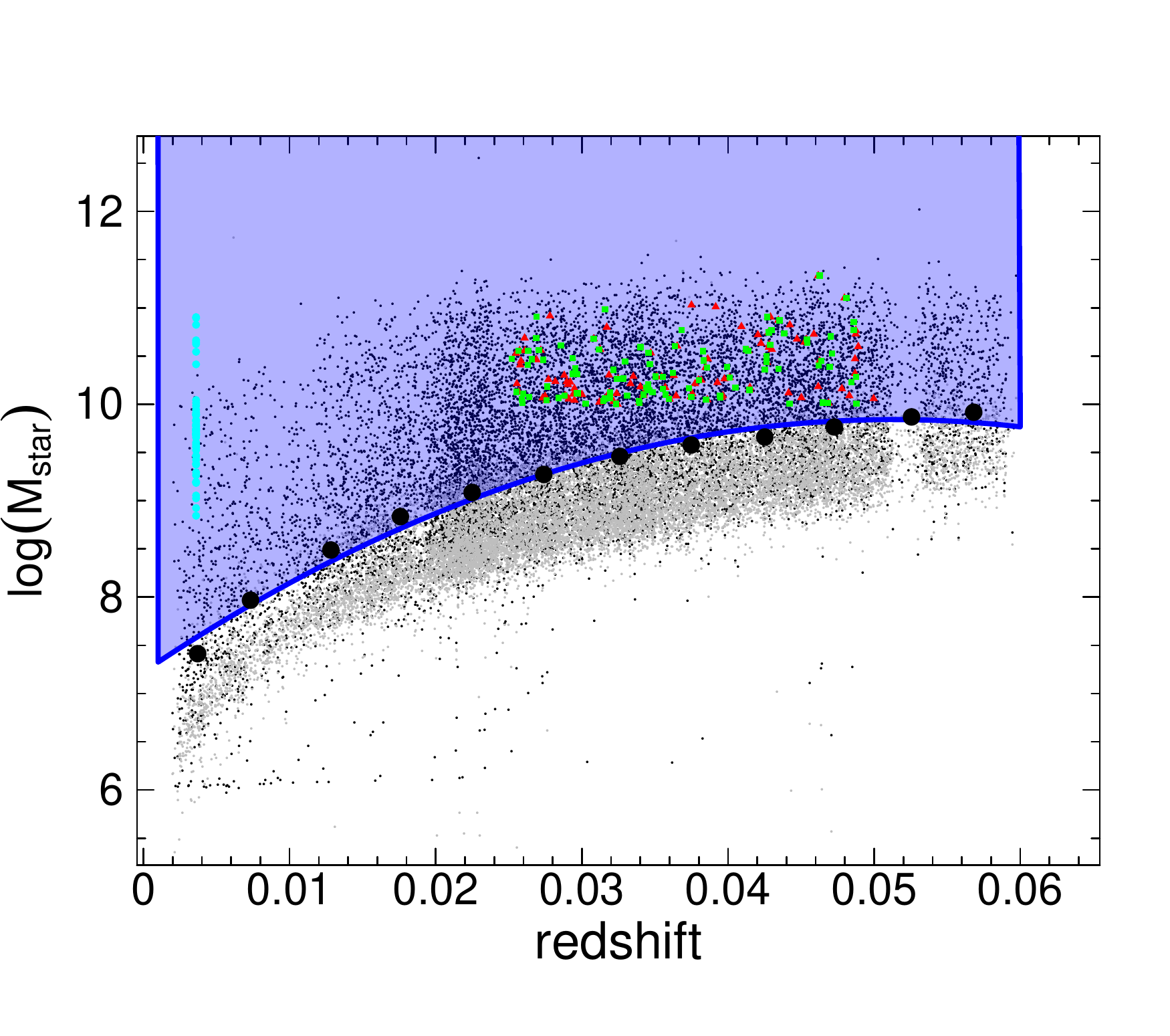}
\caption{Stellar mass as a function of redshift for our star forming sample with ALFALFA counterparts (shown in small black dots). The stellar mass each galaxy would have if its magnitude was equal to the limit SDSS magnitude is shown in gray dots ($\mathcal{M}_{lim}$). Back circles show the 95$\%$ of the M/L completeness level. Galaxies from GASS, COLDGASS and VIRGO are shown as well using the same symbols of Fig. \ref{BPT_HI} }
\label{CompSample}
\end{center}
\end{figure}

\begin{figure*}\label{Hist}
\begin{center}
\includegraphics[scale=0.5]{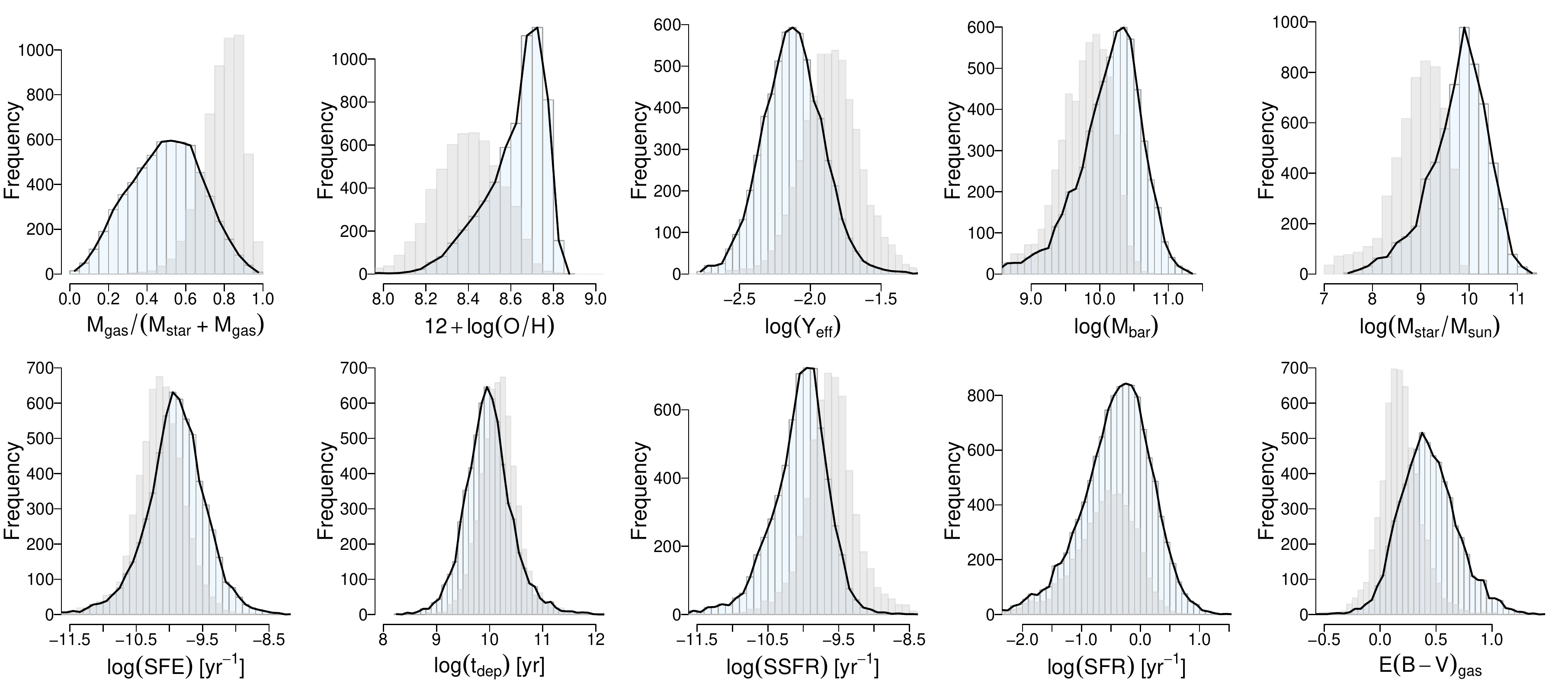}
\caption{Histograms of all our estimated variables for our final star forming sample. Light blue histograms with solid contours show our sample within the completeness limit, while light gray histograms show the sample outside the completeness limit.}
\label{Hist}
\end{center}
\end{figure*}

Since the bulk of our sample comes from SDSS and ALFALFA counterparts, we estimated a mass completeness limit to avoid biases. The main idea is to estimate the limiting stellar mass that a galaxy could have if its magnitude corresponds to a m$_{lim}$. We follow the approach by  \citet{Pozzetti10}. First, we determine for all galaxies the minimum mass $\mathcal{M}_{min}$, above which galaxies are complete. To derive $\mathcal{M}_{min}$, we calculate the limiting stelar mass $\mathcal{M}_{lim, i}$ of each galaxy, i.e., the mass it would have, at its spectroscopic redshift, if its apparent magnitude were equal to the limiting magnitude of the survey \citep[for SDSS $I_{lim}$=17.77,][]{Strauss02, Abaza09}, given by:
 
\begin{equation}\label{MasLim}
{\rm log}(\mathcal{M}_{lim, i})= {\rm log(M_{star, i})} + 0.4(I_i - I_{lim})
\end{equation}

 The result is a distribution of limiting stellar masses $\mathcal{M}_{lim}$ , that reflects the distribution of stellar M/L ratios at each redshift in our sample (gray dots in Fig.  \ref{CompSample}).  After computing $\mathcal{M}_{lim}$ for all galaxies in our sample, we selected the faintest 20$\%$ in redshifts bins of $\Delta$z = 0.005. Then, for each bin we determine the mass below which lie 95$\%$ of these faint objects, shown in black circles in Fig. \ref{CompSample}. We then define $\mathcal{M}_{min}$ as the upper envelope fit of the $\mathcal{M}_{lim}$ distribution, as shown in the same figure. For the whole paper, we will use only galaxies above the mass completeness limit (blue shaded galaxies in Fig. \ref{CompSample}). Galaxies below the mass completeness limit are shown in gray in Figs. 4-9 and 11-12, and are not used for any fit. In total, 6086 star forming galaxies lie within the completeness limit.

In summary, we have 6086 star forming galaxies with SDSS/ALFALFA counterparts within the completeness limit, 102 SDSS/GASS counterparts, 102 SDSS/COLD GASS counterparts, and 55 galaxies from the Virgo cluster. The distribution of the main properties for our final sample of SF galaxies, is shown in the  histograms of  Fig. \ref{Hist}, where blue histograms show the galaxies within the completeness limit, and grey histograms the galaxies outside the completeness limit.  From those histograms, we notice that for gas metallicities and gas fractions, the peak of the distribution is shifted considerable between galaxies within and outside the completeness limit.  Since the completeness criteria is rejecting the lowest mass galaxies at given redshift (see Fig. \ref{CompSample}), the stellar masses of the galaxies within the completeness limit are also shifted towards higher values. This also implies that galaxies outside the completeness limit will have lower stellar masses, and hence lower metallicities, and higher gas fractions.

\section[]{Scaling relationships} \label{ScalRel}

\begin{figure*}
\begin{center}
\includegraphics[scale=0.7]{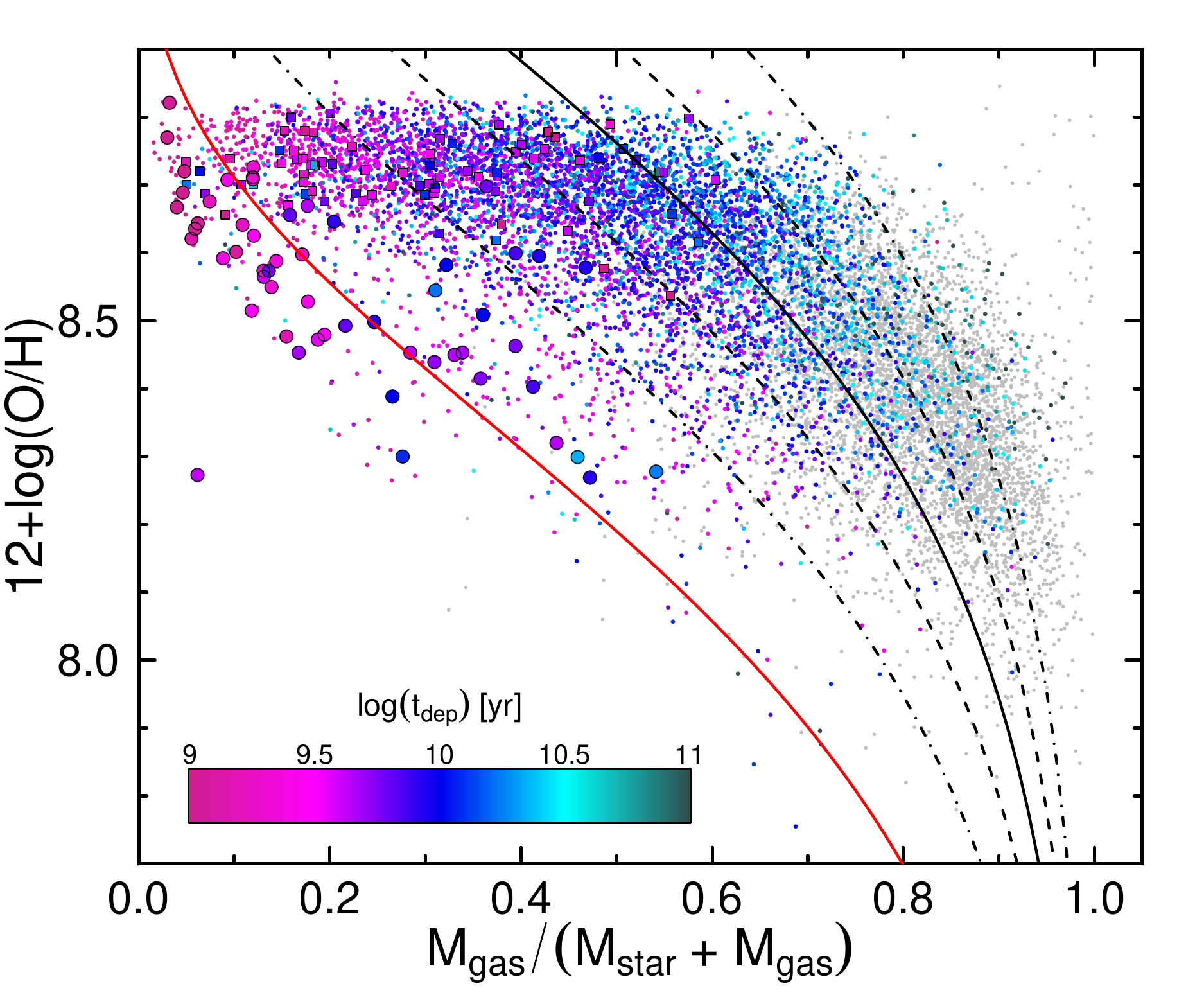}

\caption{Gas fraction and metallicity relation ($\mu$-Z) as a function of depletion time for all galaxies in our sample. SDSS/AFALFA and SDSS/GASS are shown in dots, SDSS/COLD GASS  in filled squares, and Virgo galaxies in large filled circles. Galaxies outside our completeness limit are shown in gray dots.  Using Eq. \ref{EqEffYield}, we are showing an effective yield of 0.01 in solid line, 0.014  \& 0.007 in dashed lines, and 0.0047 \& 0.021 as dot-dashed lines. The  red solid  line is the prediction of the closed-box model with oxygen yield $y_0 = 0.00268$ \citep{Pil04}.}
\label{GasFracMet}
\end{center}
\end{figure*}

In this section we present several scaling relations from the properties derived in $\S\,$\ref{SampleSelection}. First, we examine the relationship between the total oxygen abundance (estimated from Eq. \ref{MetT}), and the gas fraction ($\mu$-Z) relation, as shown in Fig.  \ref{GasFracMet}. This figure has been color coded as a function of the depletion time for all our sample. As a sanity check, we show Eq. \ref{EqEffYield} for an effective yield of  0.01 in solid line, 0.007  \& 0.014  as dashed lines, and 0.0047 \& 0.021 as dot-dashed lines. Also, we show the prediction of the closed-box model with oxygen yield $y_0 = 0.00268$ (or $-2.57$ in log units) from  \citet{Pil04} (red solid line). It is worth noting that $\sim$98 $\%$ of our SF sample has higher yields than the closed-box model prediction. 

From Fig. \ref{GasFracMet}, galaxies with low gas fraction show high gas metallicities, and are forming stars at higher efficiencies, or lower depletion times. On the other hand, galaxies with high gas fractions show low gas metallicities and longer depletion times, meaning they are inefficient at converting gas into stars. This suggests a downsizing effect in the sense that the less massive galaxies are slower in converting
their mass into stars and in building up their metal content \citep[e.g.,][]{Vale09}. Furthermore, galaxies from the Virgo cluster (shown in filled larger circles), are shifted towards lower gas fractions. As discussed by \citet[][]{Hughes13}, the most likely scenario is that the gas from these galaxies was removed, decreasing the gas fractions. Interestingly, galaxies from the Virgo cluster align very well with the closed-box model (red solid line in Fig.  \ref{GasFracMet}).  It is important to bear in mind that the $\mu$-Z relation and Oxygen yield scales will change depending on the metallicity calibration used, although the shape of this relationship should remain very similar.

\begin{figure*}
\begin{center}
\includegraphics[scale=0.3]{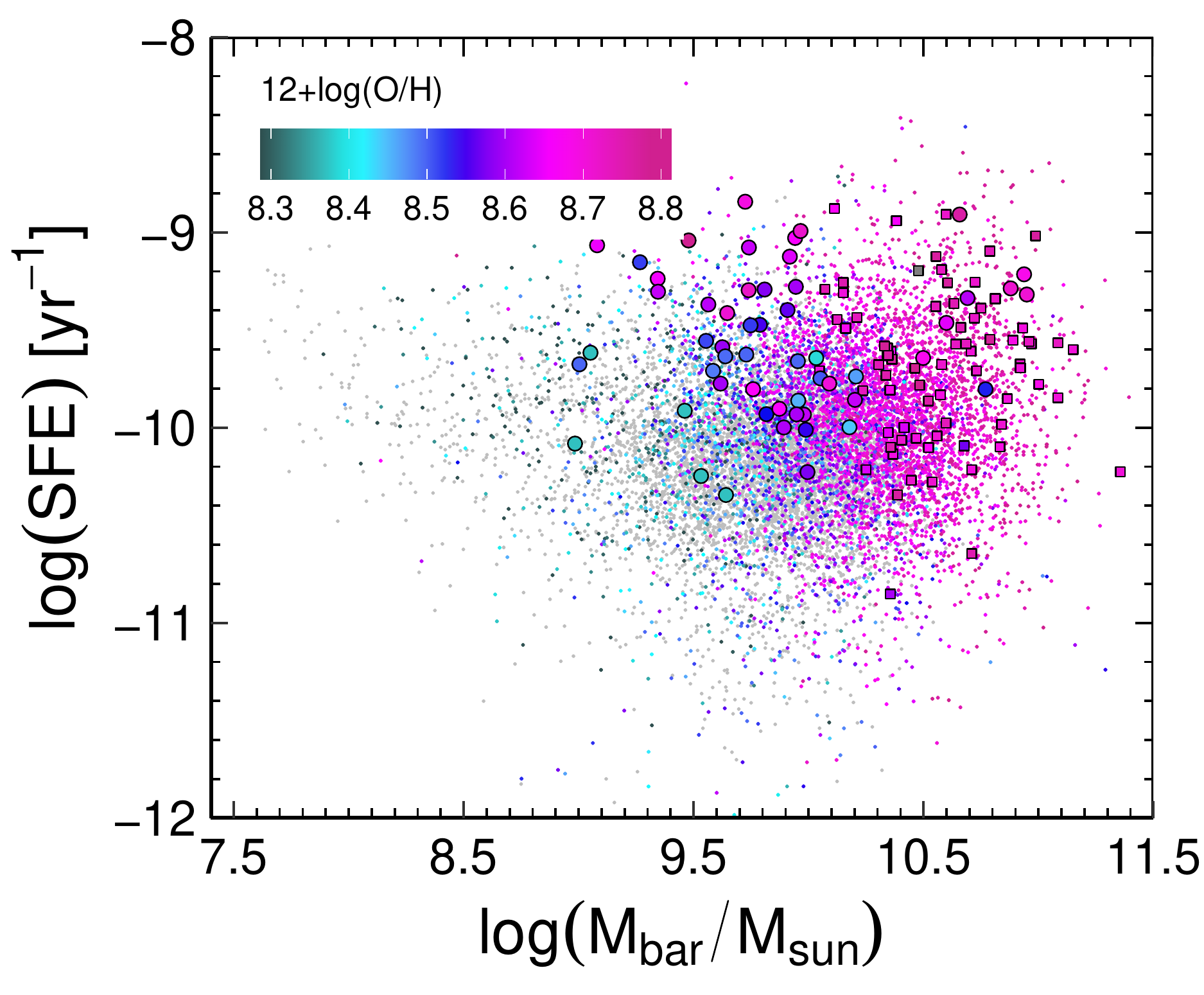}
\includegraphics[scale=0.29]{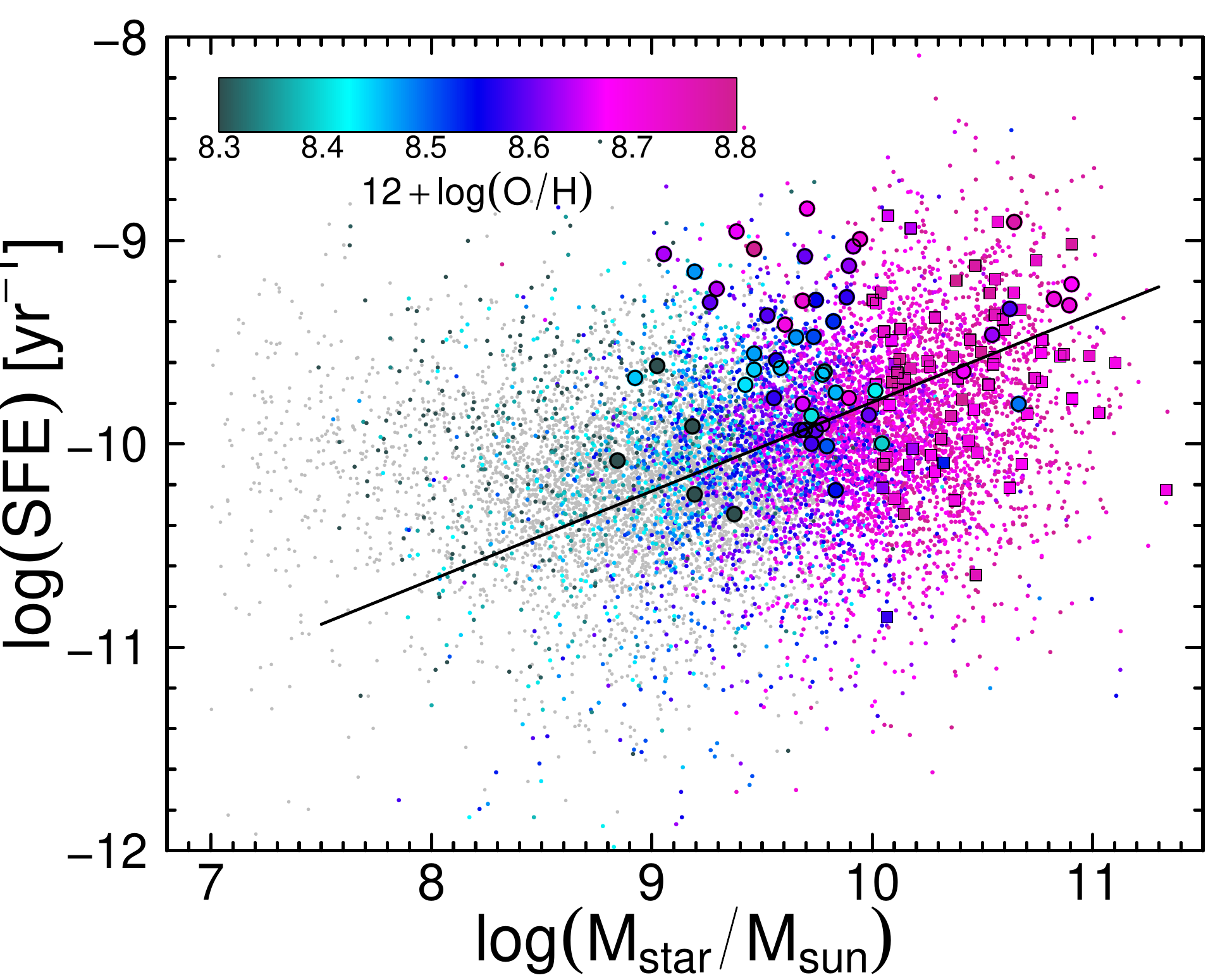}
\includegraphics[scale=0.29]{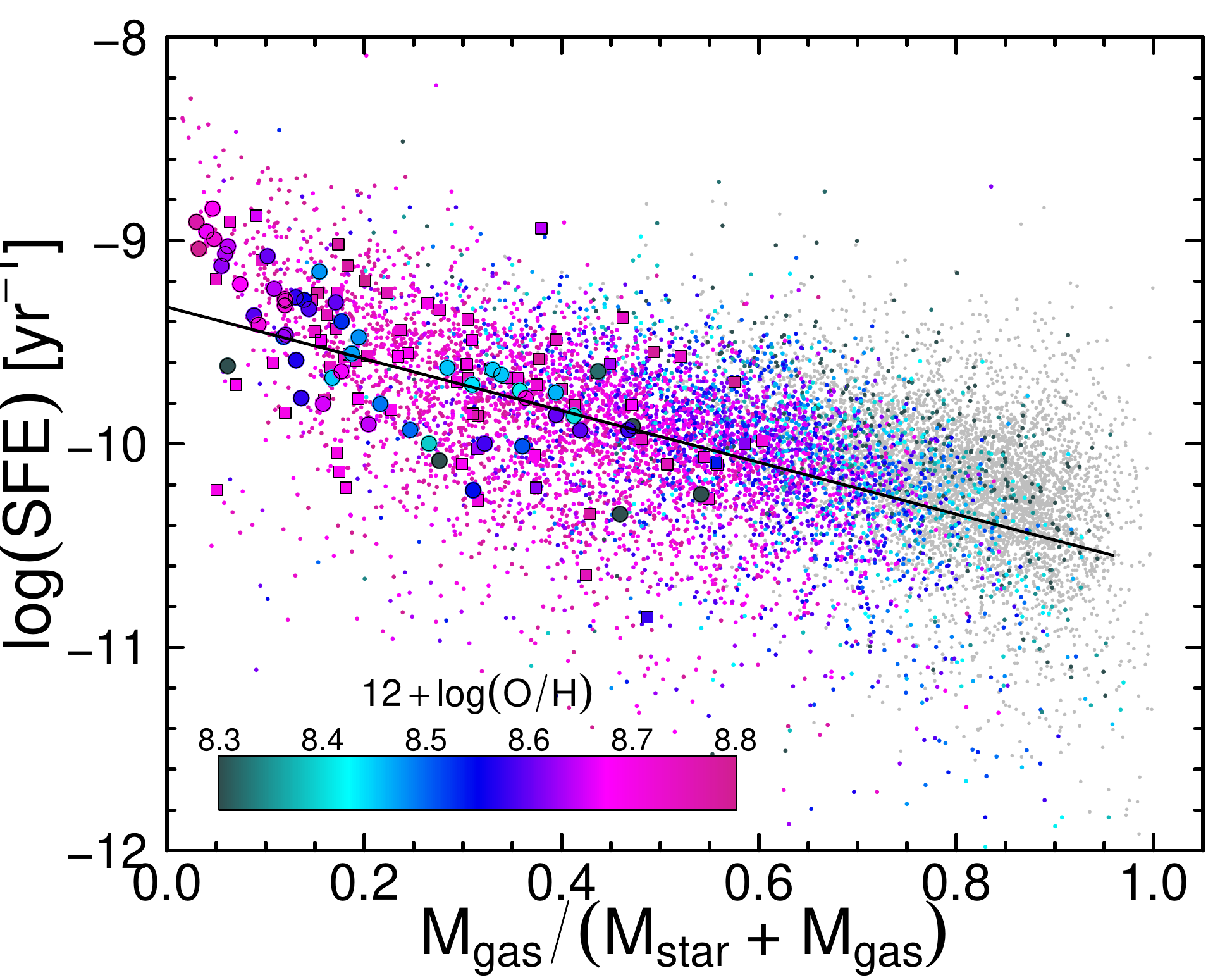}
\caption{Scaling relationships between the SFE with respect to Baryonic mass (left), stellar mass (center), and gas fraction (right). All figures are color coded as a function of the gas metallicity. Linear fits correspond only to galaxies within the completeness limit, and are given in equations \ref{MSFE} and \ref{muSFE}, respectively. Symbols are the same as in Fig. \ref{GasFracMet}. }
\label{SFERel}
\end{center}
\end{figure*}

We explore as well scaling relations with the SFE as an indicator of the current efficiency in which galaxies are converting gas into stars. We find that the SFE shows a large scatter with the baryonic mass, as shown in Fig. \ref{SFERel} (left), although suggesting opposite trends within the scatter. Galaxies with \Mb \ $>$ $\sim$ 10$^{10}$, seem to show a very high dispersion, with a slight correlation with SFE, whereas galaxies with \Mb \ $<$ $\sim$ 10$^{10}$ show a slight anti-correlation between the same variables. Furthermore, gas metallicity seems to separate one tendency from the other, as shown in the color gradient of the same figure, though we highlight that the dispersion is very high in the whole relation. 
%
%

Scaling relations involving the SFE with stellar mass or gas fraction do not suggest any bimodality, and scale with each other with a moderate dispersion for the M$_{\rm star}-$SFE relation, and  a smaller dispersion for the $\mu-$SFE relation, as shown in the center and right panels of Fig. \ref{SFERel}, respectively. The obtained fits for the M$_{\rm star}-$SFE and $\mu-$SFE relations are:

\begin{equation}\label{MSFE}
{\rm log(SFE)}=0.4365 (\pm 0.0154) \times {\rm log(M_{star})} - 14.16 (\pm 0.15)
\end{equation}

\noindent with RMSE = 0.47 

\begin{equation}\label{muSFE}
{\rm log(SFE)}= - 1.2729 (\pm 0.0242) \times {\mu } - 9.3275 (\pm 0.0128)
\end{equation}
\noindent with  RMSE = 0.37 

Linear fits are performed in ``R" with the package ``HYPERFIT" \citep{Robot15}, which uses traditional likelihood methods to estimate a best-fitting model to multidimensional data, in the presence of parameter covariances, intrinsic scatter, and heteroscedastic errors on individual data points. It assumes that both the intrinsic scatter and uncertainties on individual measurements are Gaussian, and allows for error covariance between orthogonal directions. Hereafter, all the linear fits of this paper are performed using HYPERFIT, only to those galaxies within the completeness limit, and taking into account errors in both axes.  As a goodness of fit, we estimate the  root mean square error (RMSE, defined as $\sqrt{\sum_{i=1}^{n} (y_i-\hat{y})^2 / n }$ , where $y_i$ is the $i$th observation of y, and $\hat{y}$ the predicted y value given the model).

\begin{figure}[h!]
\begin{center}
\includegraphics[scale=0.4]{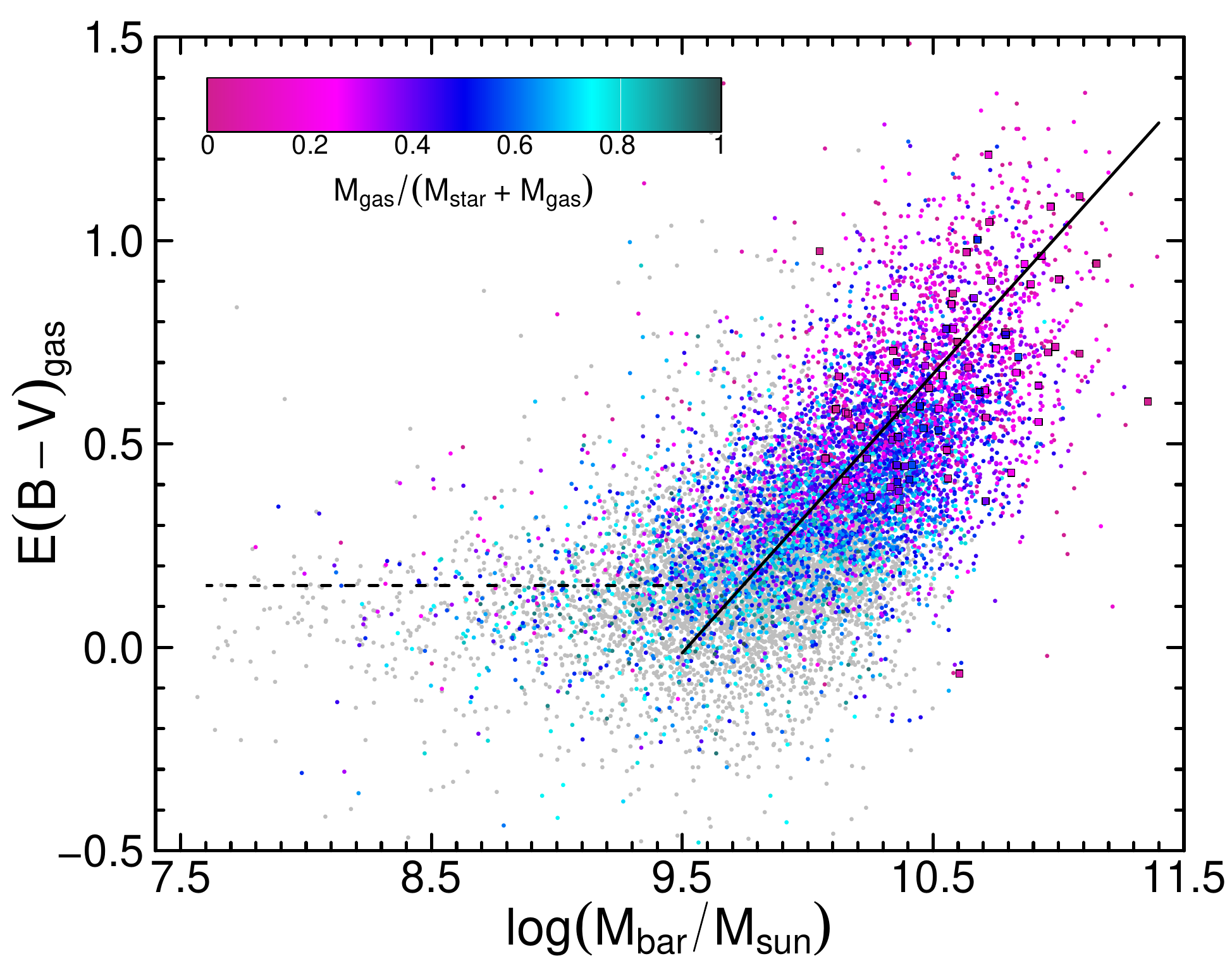}
\includegraphics[scale=0.4]{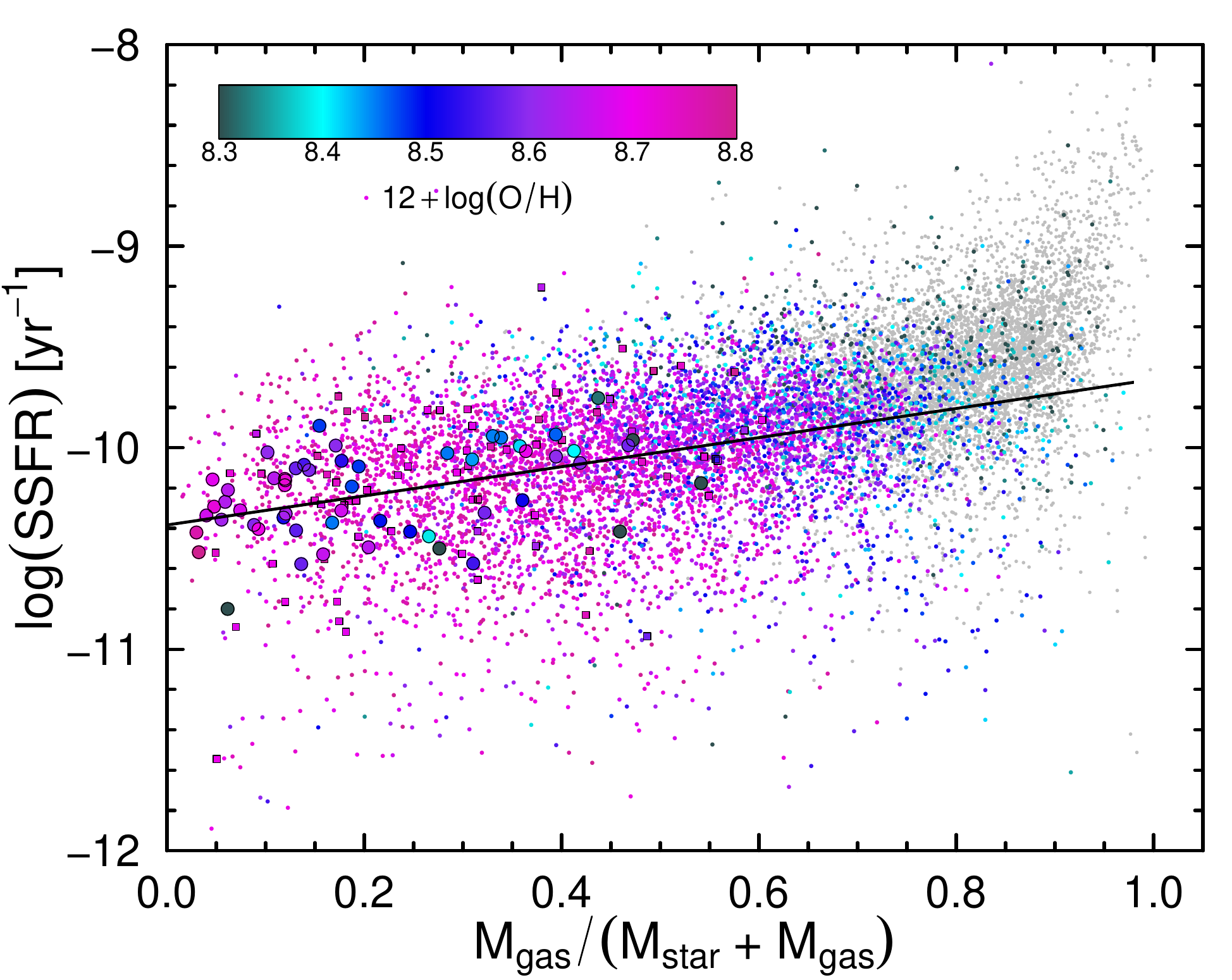}
\caption{Top: Baryonic mass vs. color excess E(B-V) as a function of gas fraction. Bottom: Gas fraction and SSFR as a function of gas metallicity. Linear fits correspond only to galaxies within the completeness limit, and are given in Eq. \ref{BmV}, and \ref{GassFracSSFR}, respectively.  . All symbols are similar as in previous figures.}
\label{DustMbar}
\end{center}
\end{figure}

Furthermore, we examine the  color excess ${\rm E(B-V)}$$_{\rm gas}$ as an indicator of extinction. We find a clear correlation of ${\rm E(B-V)}$$_{\rm gas}$ with M$_{\rm bar}$, given by Eq. \ref{BmV} where massive galaxies show a higher amount of extinction for log(\Mbar) $>$ 9.5\Msun.  We find however, that for galaxies with log(\Mbar) $<$ 9.5\Msun, the relationship tends to flatten, with a median value of ${\rm E(B-V)}$$_{\rm gas}$ = 0.15  (see Fig. \ref{DustMbar}, top). 

\begin{equation}\label{BmV}
{\rm E(B-V)_{gas}}=  0.6861 \ (\pm \ 0.0116) \times {\rm log(M}_{\rm bar}) - 6.532 \ (\pm \ 0.119)
\end{equation} \noindent with RMSE = 0.22, for  log(M$_{\rm bar}$) $>$ 9.5

On the other hand, the SSFR  increases linearly with the gas fraction (see Fig. \ref{DustMbar}, bottom), as expected due to the downsizing effect. Galaxies with very high gas fractions, are mostly outside the completeness limits, and show a very steep increase of their SSFR.  The linear relation, follows the next equation:

\begin{equation}\label{GassFracSSFR}
{\rm log(SSFR)}=  0.7228 \ (\pm \ 0.0238) \times {\mu } - 10.3838 \ (\pm \ 0.0126)
\end{equation}

\noindent with RMSE = 0.37

In general, galaxies with a high stellar mass tend to convert gas into stars more efficiently, show lower gas fractions, higher metallicities, larger dust content, and a lower SFR per unit mass, consistent with an active star formation history and an earlier assembly of their stellar mass. However, when the baryonic mass is taken as a reference, there is not a single tendency of \Mbar\ and SFE, and galaxies with a high \Mbar, show a rather scattered SFE.

On the other hand, galaxies with low stellar mass  show exactly the opposite, lower star formation efficiencies, high gas fractions,  lower metallicities, a very low dust content, and a high SFR per unit mass. However, galaxies with a low \Mbar, show a scattered anti-correlation with SFE. 

We notice a separation between low and massive galaxies only when the baryonic mass is used (Figs.  \ref{SFERel}, left, $\&$  \ref{DustMbar}, top). When gas fractions or stellar masses are used as a reference, galaxies will scale uniformly with other properties.\\

\begin{figure}[b]
\begin{center}
\includegraphics[scale=0.4]{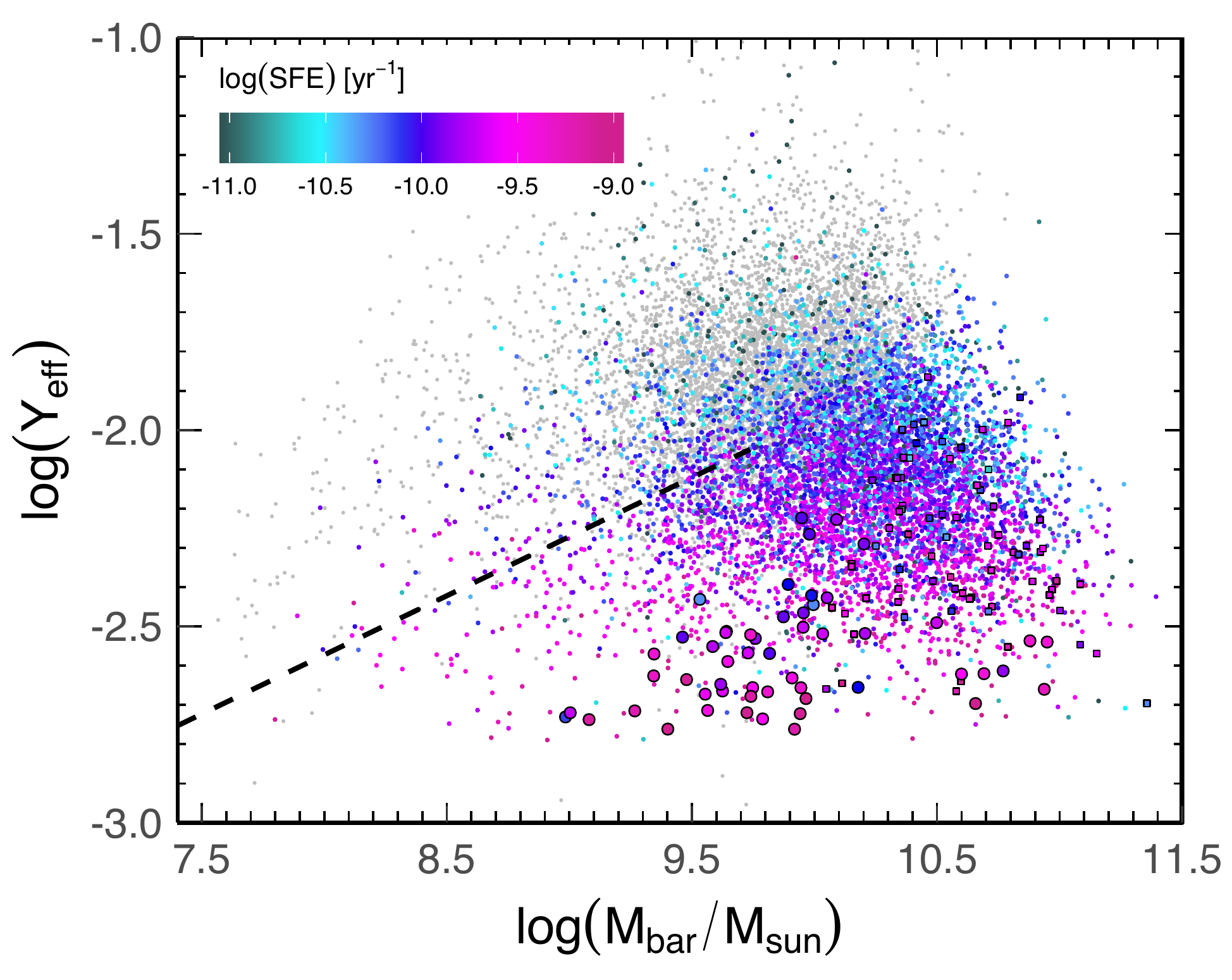}
\caption{Baryonic mass vs. Effective oxygen yields color coded with log(SFE). The dashed line shows the relationship of  \citet{Ekta10} for extremely metal-deficient galaxies (Eq. \ref{EktaEq}). All symbols are similar as in previous figures.}
\label{MbarYieldFig}
\end{center}
\end{figure}

One of the most interesting relationships we found is between M$_{\rm bar}$ and y$_{\rm eff}$, as shown in Fig. \ref{MbarYieldFig}. Besides the large dispersion shown in this relationship, we can appreciate a bimodality, when galaxies with log(\Mbar) $\lesssim$ 10\Msun show a correlation, while larger values of \Mbar\ show an anticorrelation. This bimodal behaviour is also noticeable for galaxies from the Virgo cluster  (large circles in  Fig. \ref{MbarYieldFig}). Furthermore, the y$_{\rm eff}$ seems to scale with SFE, as shown in the color code of the same figure.

This relationship was previously studied by \citet{Ekta10} for low metallicity galaxies. They analyzed a set of extremely metal-deficient (XMD) galaxies, blue compact galaxies, and dwarf irregular galaxies, with metallicities $\leq$ 1/10 solar. They find that the y$_{\rm eff}$ increases with baryonic mass in their Fig. 4 (bottom), which is valid from 7 $<$ log(\Mbar) $<$ 9.7,  the corresponding equation to their relation is given by: 

\begin{equation}\label{EktaEq}
{\rm log(y}_{\rm eff})_{\rm Ekta} = 0.30116 \times {\rm log(M_{bar})} - 4.9818
\end{equation}

In  Fig. \ref{MbarYieldFig}, we show the fit of \citet{Ekta10} in dashed lines, finding a very good agreement with our low mass sample. This strengthen the idea that for log(\Mbar) $\lesssim$ 10\Msun, there is a correlation between  M$_{\rm bar}$ and y$_{\rm eff}$. Furthermore, other authors have agreed that for low mass galaxies, y$_{\rm eff}$ correlates with M$_{\rm bar}$ \citep[e.g.,][]{Garnett02, Tremonti04,Lee06}.

We note that \citet{derossi2017} reported a similar behaviour for the  $M_{\rm bar}-y_{\rm eff}$ relation at $z=0$
when analysing EAGLE simulations (see their Fig.7).  However, in general, these authors obtained lower values
of $y_{\rm eff}$, which might be a consequence of a different metallicity calibration, as well as the larger aperture they used to estimate metallicities (30 kpc). Also, in that work, gas fractions were estimated considering only
star-forming gas particles inside 30 kpc, while here, we consider the whole gas component within
70 kpc. See Appendix \ref{Ap2} for more details about aperture effects on the $M_{\rm bar}-y_{\rm eff}$ relation
in EAGLE simulations.


 From the oxygen yields histogram in Fig. \ref{Hist} and Fig. \ref{MbarYieldFig}, we find that the highest values ( log(y$_{\rm eff}$) $> -2$ ) correspond to galaxies with a similar metallicity distribution shown in  Fig. \ref{Hist}, but higher gas fractions, with a median of $\sim$0.7, almost 0.2 dex higher than the general gas fraction distribution in Fig. \ref{Hist}. It is been suggested by \citet{Vincenzo16} that high values of effective yield may be indicative of an IMF favouring massive stars \citep[see also][]{Molla15}.

In order to understand the  suggested bimodality of the M$_{\rm bar}$ $-$ y$_{\rm eff}$ relation in Fig. \ref{MbarYieldFig}, we will now examine the stellar ages of our galaxies and perform a more detailed analysis of this relationship with more dimensions ($\S\,$\ref{Discussion}). Furthermore, we seek different scenarios and the feedback processes that can produce such a bimodality by investigating the results from the EAGLE simulations ($\S\,$\ref{EAGLEsim}).

\subsection[]{Stellar Ages} \label{Age}

The stellar age in galaxies is key to understand the timescales in which galaxies have converted their gas into stars.  We use the Bayesian-likelihood estimates of the r-band light-weighted ages (t$_{\rm r}$) for the SDSS-DR7 estimated by \citet{Gallazzi05}. In summary, the stellar ages are derived by comparing the spectrum of each galaxy to a library of  \citet{Bruzual03}  models at medium-high spectral resolution, encompassing the full range of physically plausible star formation histories. Similarly to \citet{Kauf03a}, they adopted a Bayesian approach to derive a posteriori likelihood distribution of each physical parameter by computing the goodness of fit of the observed spectrum for all the models in the library.

\begin{figure*}[h]
\begin{center}
\includegraphics[scale=0.3]{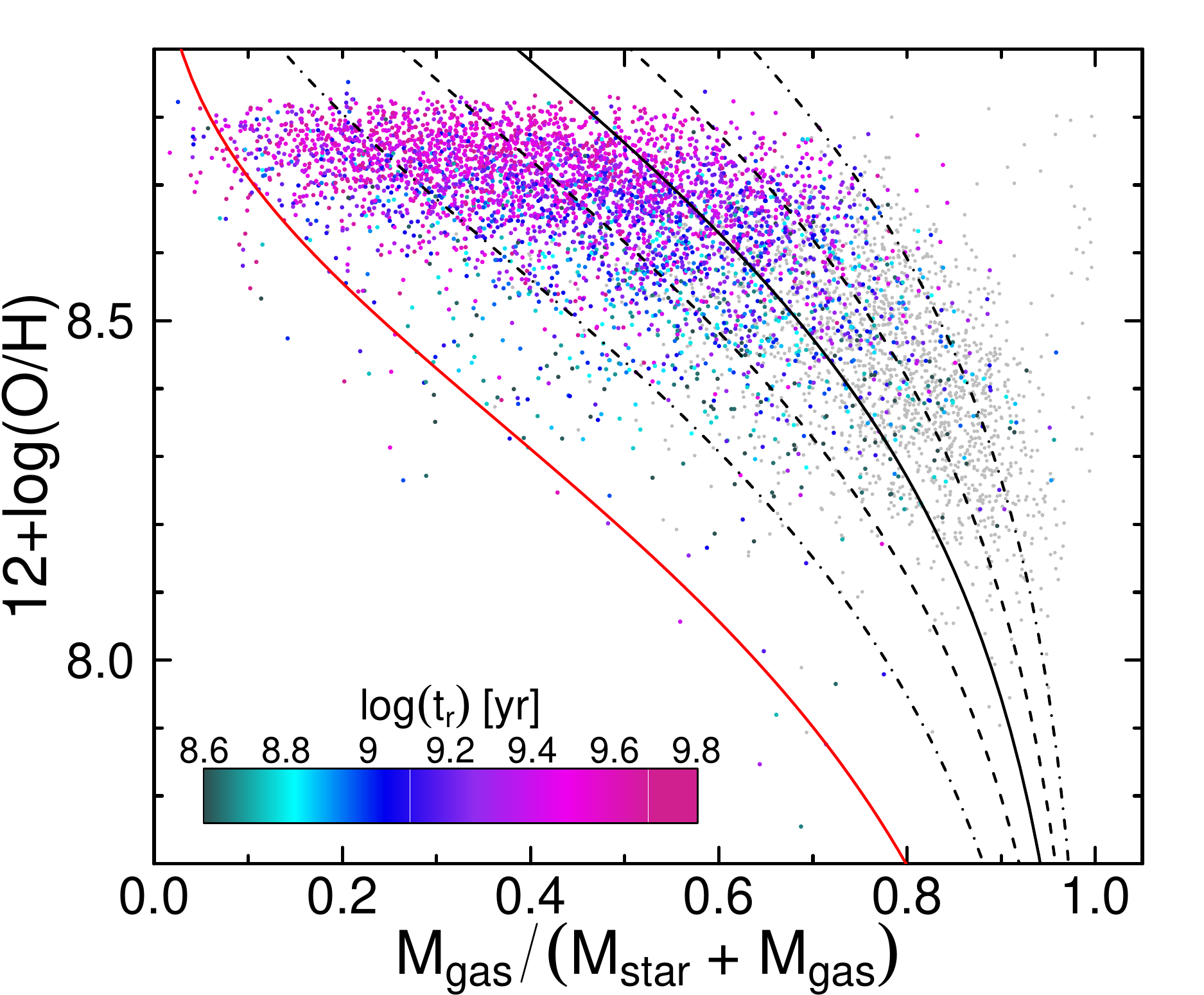}
\includegraphics[scale=0.3]{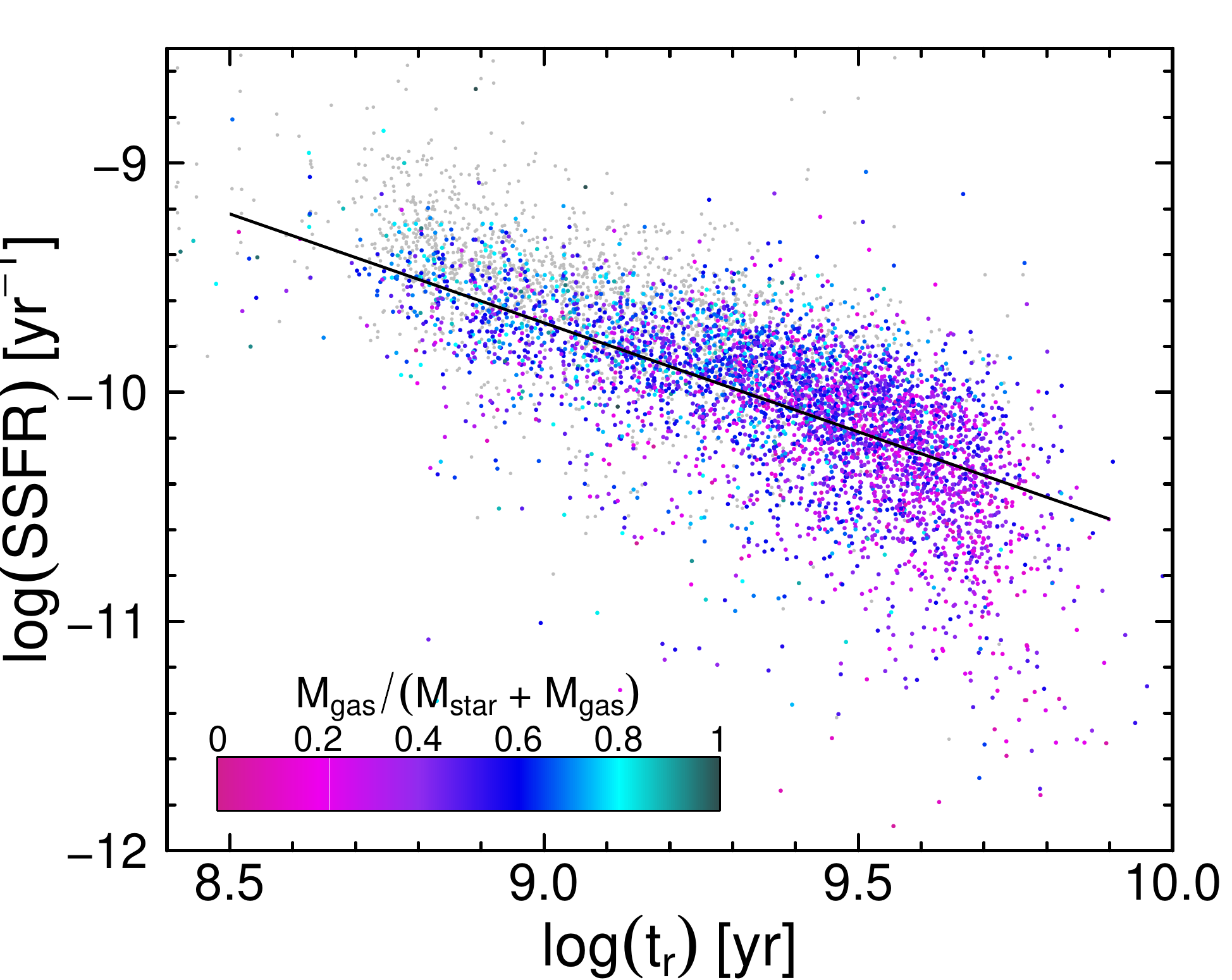}
\includegraphics[scale=0.3]{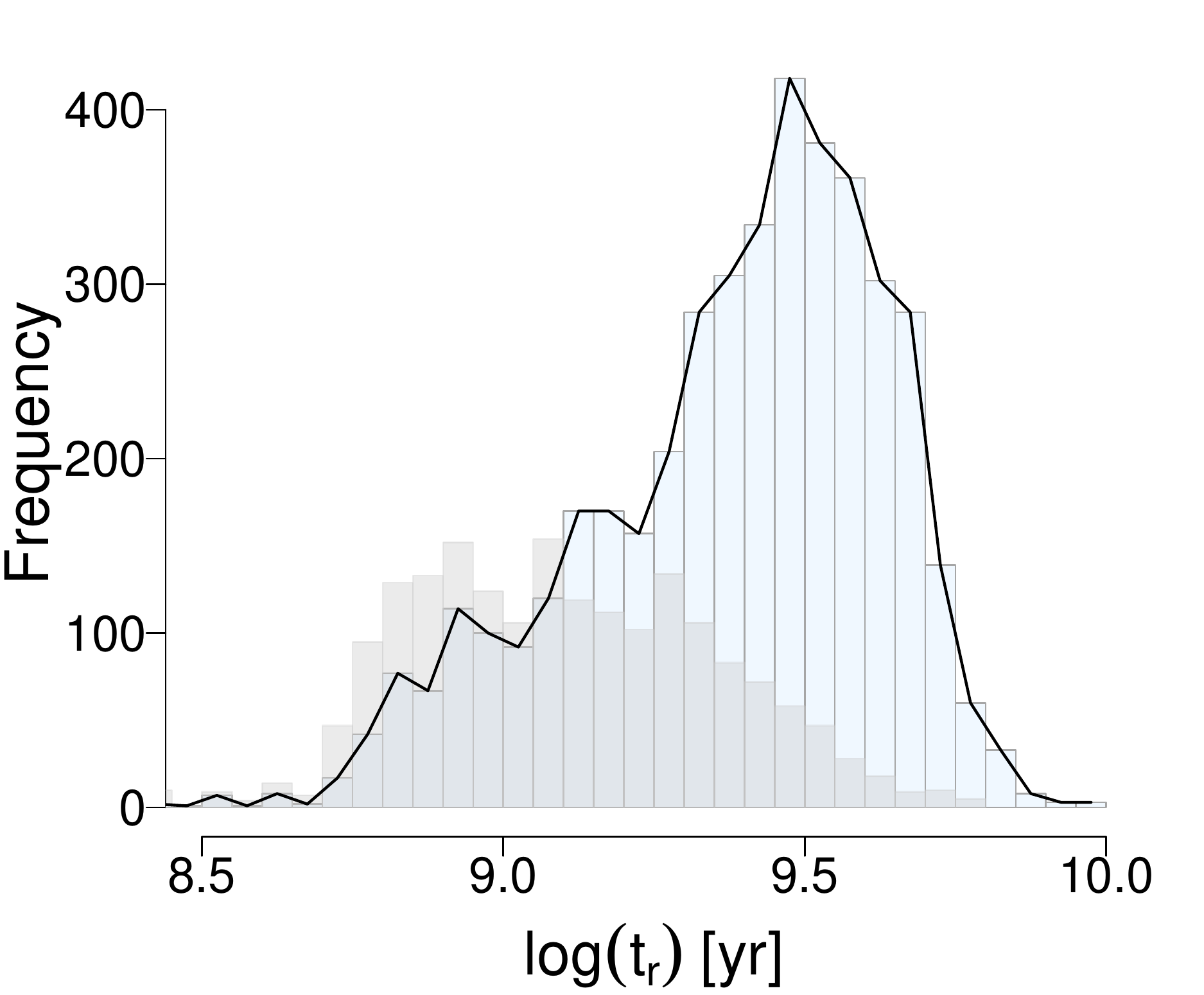}
\caption{Left: $\mu$-Z relation as a function of age weighted by r-luminosity (t$_{\rm r}$), Center: t$_{\rm r}$-SSFR relation as a function of gas fraction. Right: Ages histogram. Galaxies within the completeness limit are color coded (4266 galaxies), while galaxies outside the completeness limit with an stellar age measured are shown in grey  (1888 galaxies). }
\label{AgePlots}
\end{center}
\end{figure*}

After matching the stellar ages catalog with the SDSS/ALFALFA SF sample, we obtained 4266 galaxies within the completeness limit and 1888 galaxies outside the completeness limit. The stellar ages are distributed from $\sim$ 10$^{8.6}$ - 10$^{9.9}$ years (see Fig \ref{AgePlots}, right), with a negative skew that could be indicative of a second, younger population. We show the $\mu$-Z relation as a function of stellar age (Fig \ref{AgePlots}, left), as appreciated, a large fraction of the galaxies have high gas metallicities. Additionally, we find that the t$_{\rm r}$-SSFR relation is tightly anti-correlated, and have included this relationship as it may be useful to have a rough estimate of the stellar age when it cannot be estimated through more sophisticated methods. The obtained fit is:

\begin{equation}\label{AgeSSFR}
{\rm log(SSFR)}= - 0.9509 \ (\pm \ 0.0156) \times {\rm  log(t_r) } - 1.1391 \ (\pm \ 0.1462)
\end{equation}

\noindent with  RMSE = 0.29

A further discussion on the implication of the stellar age is given in $\S\,\ref{Discussion}$.

\section[]{Confronting simulations with observations}\label{EAGLEsim}

In order to get some insight into the astrophysical processes which could drive the
observed scaling relations for $y_{\rm eff}$, we compared 
our results with predictions
from the {\sc EAGLE} suite of cosmological simulations \citep[][]{crain2015, schaye2015}.\footnote{See
http://eagle.strw.leidenuniv.nl and http://www.eaglesim.org/ for different data
products, images and movies \citep{mcalpine2015, eagle2017}.}  
In Sec.~\ref{sec_eagle_description}, we briefly describe the set of {\sc EAGLE}
simulations used in this work and the simulated galaxy sample.
A comparison between observational data and EAGLE results is presented in 
Sec.~\ref{sec_eagle_comparison}, while in Sec.~\ref{sec_eagle_AGN} we discuss
the effect of AGN feedback on effective yields.

\subsection{{\sc EAGLE} simulations}
\label{sec_eagle_description}
{\sc EAGLE} simulations were performed by using a modified version of the {\sc GADGET-3} hydrodynamical
code \citep{springel2005}. These simulations track the joint evolution of dark matter and baryons within
different periodic comoving volumes, adopting different resolutions. The cosmology used is by EAGLE is $H_0=67.77\,$km\,s$^{-1}$\,Mpc$^{-1}$, $\Omega_M=0.307$, and $\Omega_{\Lambda}=0.693$.  Different processes are included in the {\sc EAGLE}
subgrid physical model, such as radiative cooling and heating of the gas, 
star formation, chemical enrichment, stellar and AGN feedback, among others (see \citealt{schaye2015, crain2015},
for more details).

Here, we will focused on the analysis of the so-called Recal-L025N0752,
Ref-L050N0752, AGNdT9-L050N0752, and NO-AGN-L050N0752 simulations, as they provide important
clues for the understanding of the observed $y_{\rm eff}$ trends (see below).
The suffix in the name of the simulations indicate the box length in comoving megaparsec (cMpc, e.g. L025)
and the cube root of the initial number of particles per species (e.g. N0752).
For simulations run with the reference model (denoted with the prefix `Ref'), the
subgrid parameters associated to energy feedback were calibrated to obtain good 
agreement with the $z = 0.1$ galaxy stellar mass function (GSMF), 
whilst also reproducing the observed sizes of present-day disc galaxies.
Higher resolution simulations (e.g. Recal-L025N0752) implement also a `Recal-' model, which uses
subgrid parameters that have been recalibrated to improve the fit to the $z\sim0$ GSMF
when increasing the resolution. Variations of AGN feedback parameters have only been 
explored within comoving volumes of side lengths of 50 cMpc ( `L050' runs); thus, the
latter simulations will be used to study the impact of AGN feedback on $y_{\rm eff}$.
In particular, simulation AGNdT9-L050N0752 assumes a temperature increment associated to AGN
heating of $\Delta T_{\rm AGN}=10^9 \ {\rm K}$ while, for the reference model, 
$\Delta T_{\rm AGN}=10^{8.5} \ {\rm K}$. Increasing $\Delta T_{\rm AGN}$ drives more energetic
individual feedback events, generally leading to smaller radiative losses in the ISM.  Larger
values of $\Delta T_{\rm AGN}$ produce a more intermittent feedback process (see \citealt{derossi2017}
for a deep analysis of AGN feedback effects on the global chemical enrichment of galaxies).
In the case of the simulation NO-AGN-L050N0752, the black hole (BH) model is turned off:
BH gas accretion and AGN feedback are completely disabled.

Throughout  the current work, the global properties of simulated galaxies that are typically 
measured from optical diagnostics (e.g. M$_{\rm star}$, SFRs, stellar ages, chemical abundances), 
are estimated considering bound particles within a spherical aperture of radius 30 kpc centred 
on the potential minimum of the system.  According to \citet{schaye2015}, 
stellar masses derived in this way are comparable to those obtained within a projected
circular aperture of the Petrosian radius. 
As oxygen abundances are typically inferred from SF gas HII regions, 
 in this work, we evaluate chemical abundances taking into account only {\em star-forming} gas particles,
similarly to the procedure applied by \citet{derossi2017}. In Appendix \ref{Ap2}, we show that the main
trends found for EAGLE simulated yields are robust against aperture effects.


For estimating gas masses (M$_{\rm gas}$), we follow \citet{crain2016} and 
consider a larger aperture of 70 kpc,
which roughly corresponds to the Arecibo L-Band Feed Array (ALFA) FWHM beam size 
of $\sim3.5$ arcmin \citep{giovanelli2005} at the median redshift of the GASS sample, $z = 0.037$
\citep{Catinella10}. In the case of simulations, we define M$_{\rm gas}$ as the total
hydrogen mass enclosed by 70 kpc. We have checked that, for the scaling relations studied here, 
similar general trends are obtained if defining M$_{\rm gas}$ as the neutral
hydrogen mass (M$_{\rm Hn}$), where M$_{\rm Hn}$ was calculated following previous works
\citep[e.g.][]{marasco2016}.  As the estimate of M$_{\rm Hn}$ involves further assumptions and 
approximations, for the sake of simplicity, we do not applied such separation in this work.
 The larger aperture of 70 kpc used for estimating $M_{\rm gas}$ leads to $\approx 0.1$
higher gas fractions than those obtained if using an aperture of 30 kpc.  As we will shown
in Appendix \ref{Ap2}, if we measure all simulated quantities within 30 kpc, only moderate changes
are obtained in the normalization and scatter of the $M_{\rm bar}-y_{\rm eff}$ relation
but the main trends are preserved.

Finally, to avoid resolution issues, in our analysis we only consider the simulated galaxies with
M$_{\rm star} \ge 10^9 \ {\rm M}_{\sun}$ \citep{schaye2015}.
Unless otherwise specified, we only present results for galaxies with M$_{\rm star} \ge 10^9 \ {\rm M}_{\sun}$.

\subsection{Comparison with observations}
\label{sec_eagle_comparison}

Within the {\sc EAGLE} suite of simulations, the high-resolution Recal-L025N0752 run shows 
the best agreement with the observed metallicity scaling relations
\citep[e.g.][]{schaye2015, derossi2017, derossi2018}. 
In particular, \citet{derossi2017} found that these simulations are able to 
reproduce the observed trends for the mass-metallicity relation and its secondary dependences 
on gas fraction, SFR, SSFR, and stellar age.
Regarding effective yields, the Recal-L025N0752 run also predicts effective yields 
consistent with observations of local low-mass galaxies (see \citealt{derossi2017}, Fig. 7);
but, at higher masses, $y_{\rm eff}$ tends to decrease with mass, departing
from the observed behaviour reported by \citet{Tremonti04}. It is worth noting
that \citet{derossi2017} estimated gas masses considering SF gas particles within
30 kpc and such definition is different to that applied here (Sec.~\ref{sec_eagle_description}).
So, caution should be taken when comparing our results with those obtained by \citet{derossi2017}.
\citet{Tremonti04} derived gas masses indirectly from the observed SFRs assuming
a Schmidt law \citep{kennicutt1998}.
We will show that effective yields obtained from the Recal-L025N0752 simulations and observational
results presented here are in better agreement over a larger mass range.

\begin{figure}[h]
\begin{center}
\includegraphics[scale=0.29]{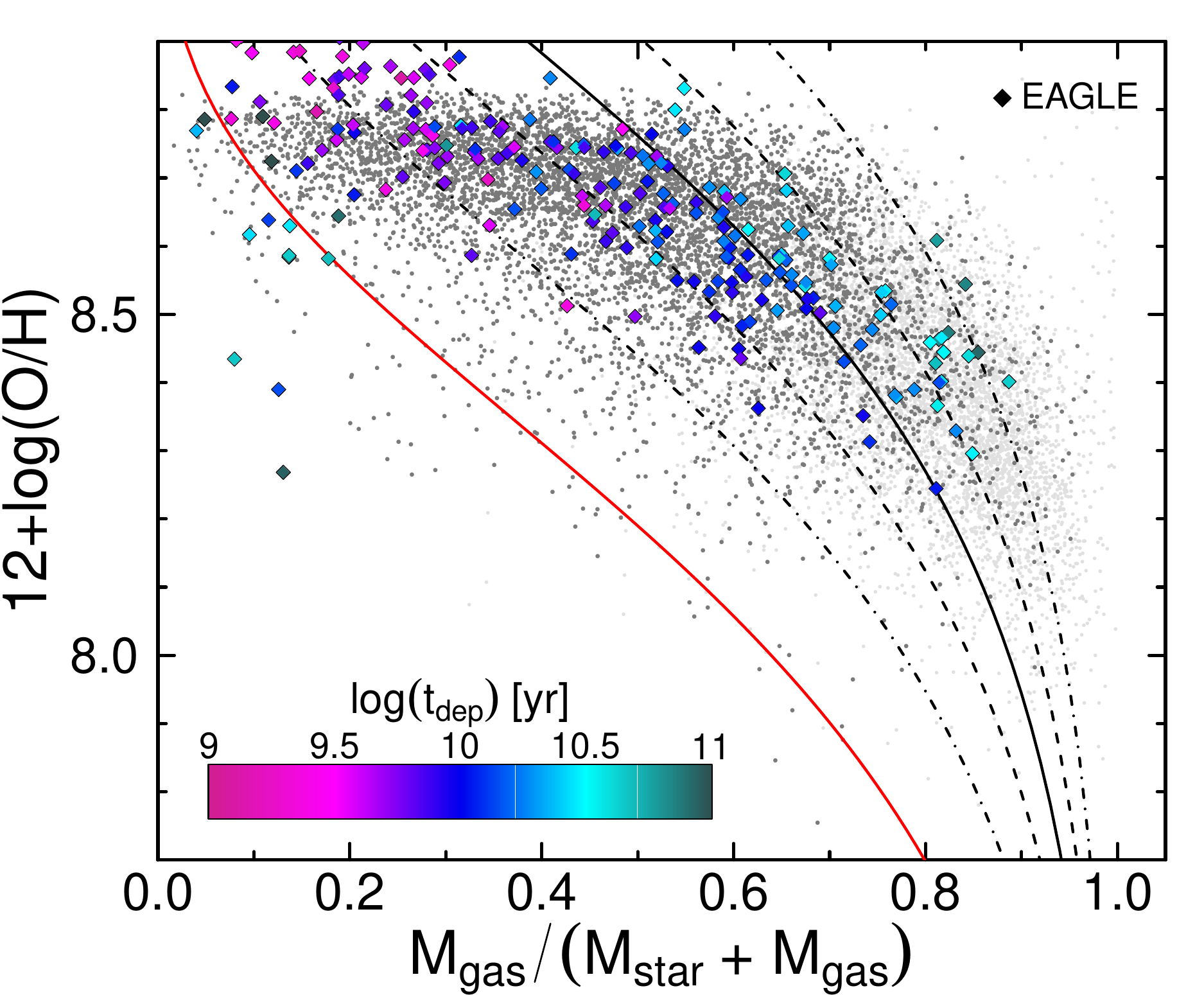}
\includegraphics[scale=0.29]{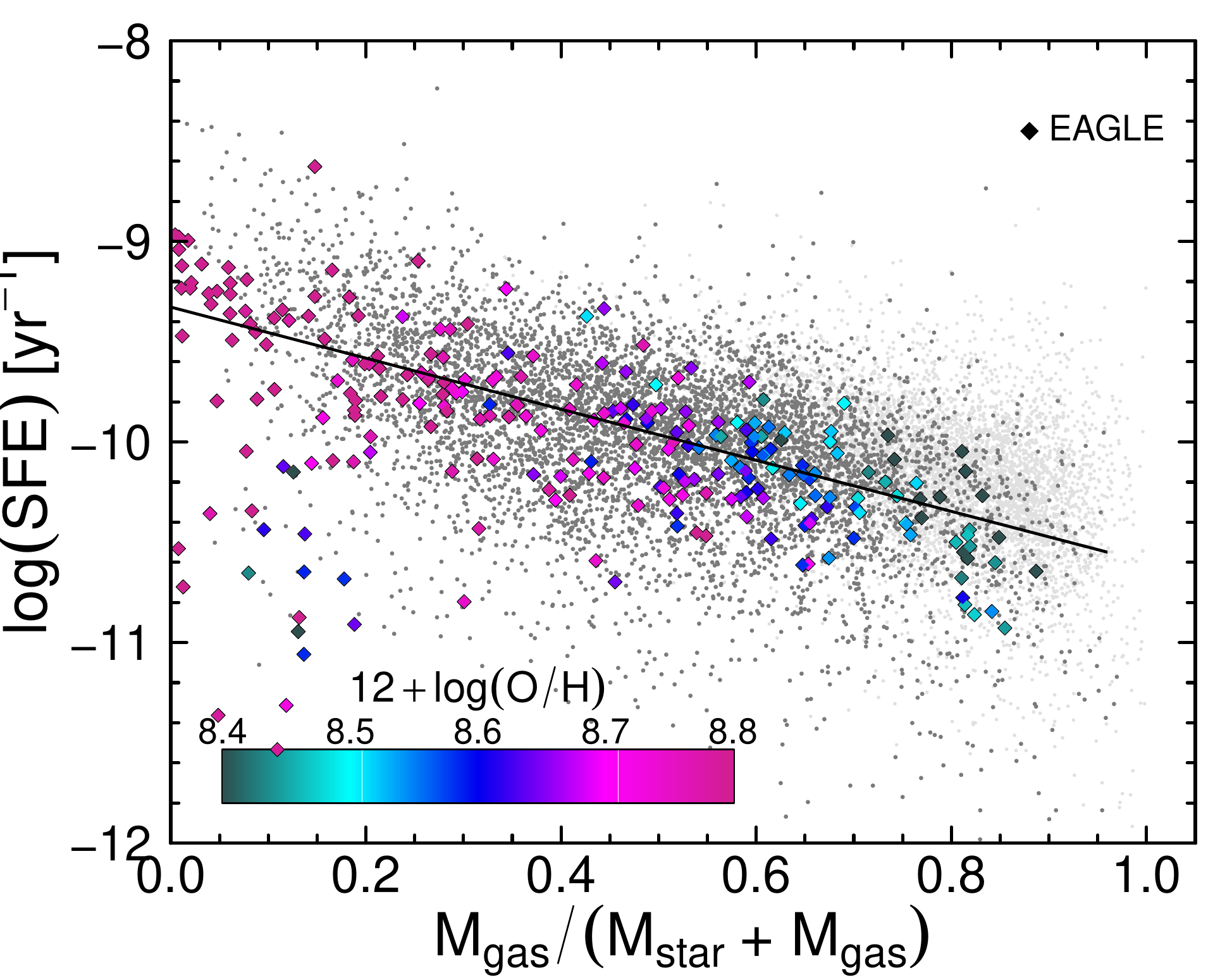}
\includegraphics[scale=0.29]{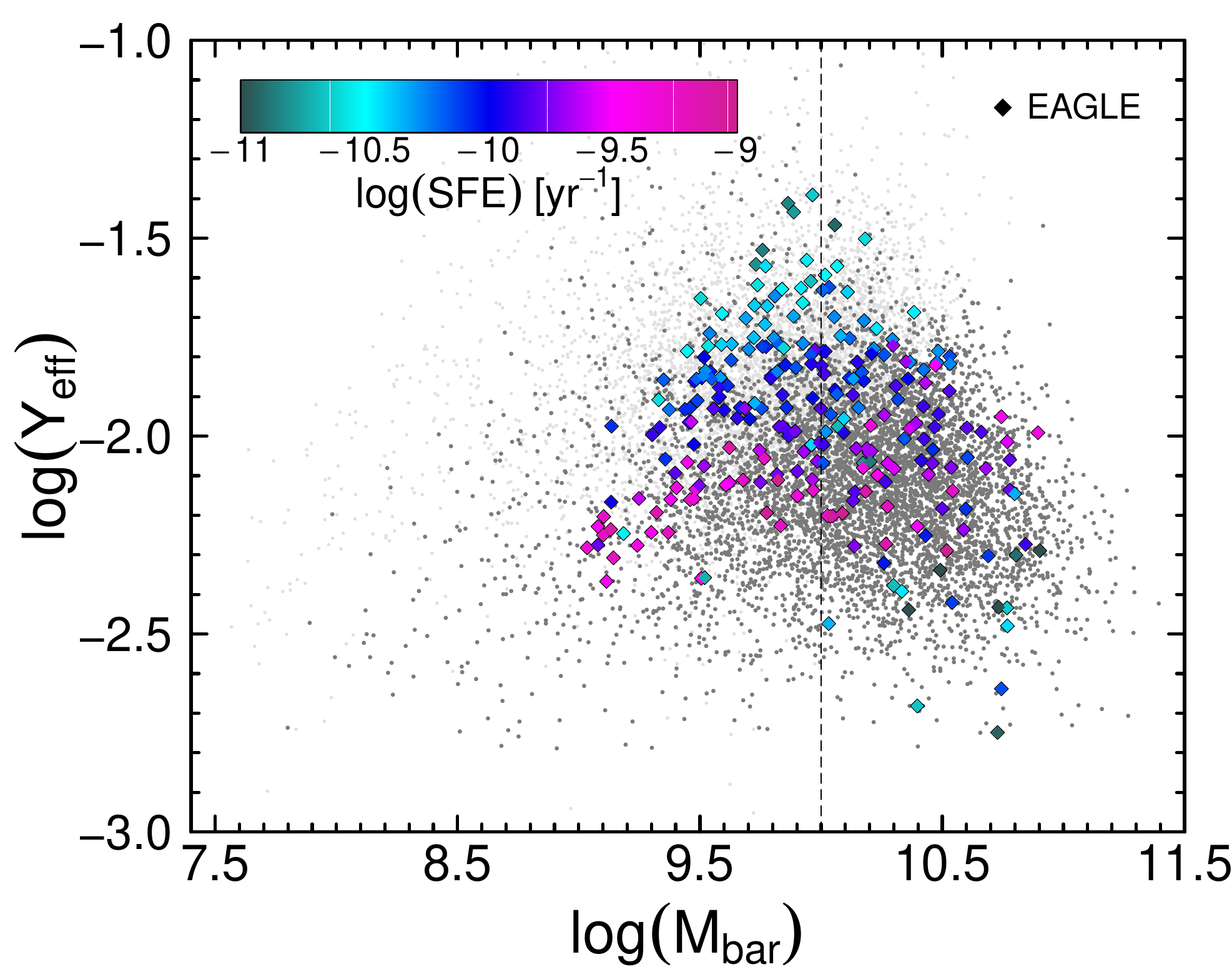}
\includegraphics[scale=0.29]{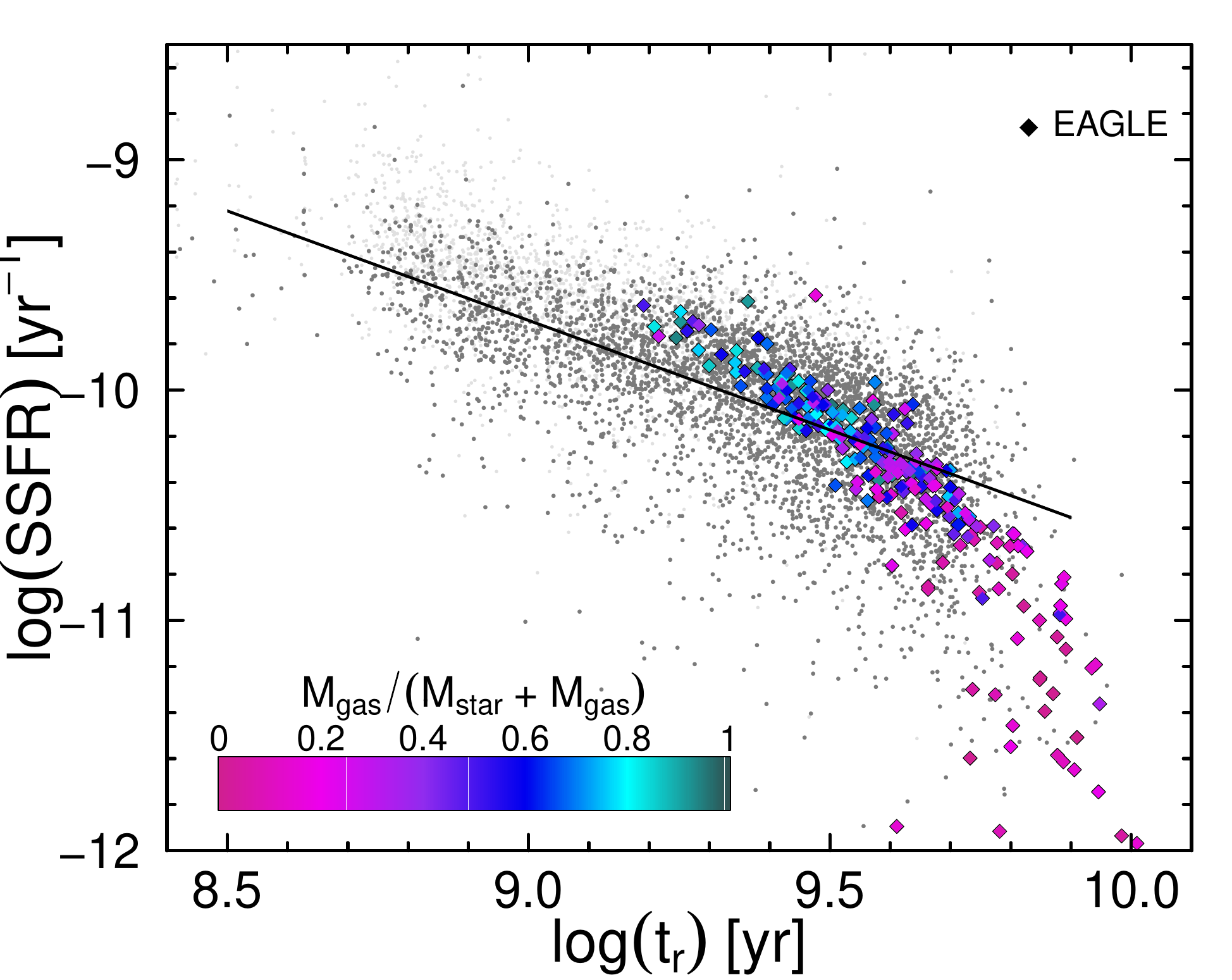}
	\caption{From top to bottom, $\mu$-Z, $\mu$-SFE, M$_{\rm bar}$-y$_{\rm eff}$, and t$_{\rm r}$-SSFR relations. {\sc EAGLE} high resolution simulations corresponding to Recal-L025N0752 are shown in diamonds color coded as the variable in the inset color bar. Dark-gray dots show our main ALFALFA/SDSS sample within the completeness limit, while light-gray dots show galaxies outside the completeness limit. The fits shown in solid lines for the 2nd and 4th panels, from top to  bottom, correspond to Eq. \ref{muSFE}, and \ref{AgeSSFR}, respectively. }
\label{EagleVsObs}
\end{center}
\end{figure}

As a sanity check, in Fig. \ref{EagleVsObs} we show a direct comparison between our main ALFALFA/SDSS sample, and the result of the high resolution {\sc EAGLE} simulations. Simulations and observations are in a perfect agreement. Furthermore, even the dispersion in the scaling relations is recovered, where the M$_{\rm bar}-$y$_{\rm eff}$ relation shows a bi-modal behaviour, and the t$_{\rm r}-$SSFR relation shows a very tight anti-correlation.  We only note that {\sc EAGLE} 
galaxies tend to be older than observed ones, which might be a consequence of the lower mass cut that we imposed to our simulated sample 
(M$_{\rm star} \ge 10^9 \ {\rm M}_{\sun}$).

 In order that simulated galaxies are well resolved and to avoid numerical artifacts, we follow De Rossi et al. (2017) and
analyse only galaxies with M$_{\rm star}$ $>$ $10^9$ \ ${\rm M}_{\sun}$ \citep[see also][]{schaye2015}.  With this mass limit, we
obtain, for example, more than 7000 baryonic particles within an aperture of 30 kpc for the Recal-L025N0752
simulation. We note, however, that this lower mass cut is below the mass limit associated to the observed sample
within the completeness limit ($> 10^7$ ${\rm M}_{\sun}$, Fig. \ref{CompSample}).  Had we included in our analysis the sub-sample of simulated
galaxies with $10^7$  ${\rm M}_{\sun}$ $<$ M$_{\rm star} < 10^9 {\rm M}_{\sun}$, higher effective yields would have been obtained at the low-mass end of our analysed scaling relations, reaching log(y$_{\rm eff}$) $\sim$ $-1$ in some cases.
The increase of  log(y$_{\rm eff}$)  is explained considering that galaxies within such low mass range, tend to show higher than average gas fractions, SSFRs and lower than average SFEs and ages (see Fig. 10). Finally, we note that a mass limit of $10^7$ ${\rm M}_{\sun}$ selects systems poorly resolved (less than 50 baryonic particles within 30 kpc) even for the Recal-L025N0752 simulation (the higher resolution simulation
within the EAGLE suite). Therefore, we decided to apply a cut at a higher mass ($10^9$ ${\rm M}_{\sun}$ ).

We also notice that high values of $y_{\rm eff}\ga -1.8$ (over solar) are predicted
for simulated gas-rich galaxies with the lowest SFE, highest SSFR and youngest ages.

\subsection{Impact of AGN feedback}
\label{sec_eagle_AGN}

The third panel in Fig.~\ref{EagleVsObs} shows that observations and simulations predict a 
correlation between $y_{\rm eff}$ and M$_{\rm bar}$ at low masses.  This behavior has been previously
explained in the literature by invoking the higher efficiency of galactic winds to eject metals from the shallower
potential wells of low-mass galaxies. 
On the other hand, at high masses, a flattening of the  \Mbar$-y_{\rm eff}$ relation was reported by 
\citet{Tremonti04}.  Interestingly, our main ALFALFA/SDSS and {\sc EAGLE} samples predict an
anti-correlation between $y_{\rm eff}$ and M$_{\rm bar}$ for massive galaxies. It is important to note that \citet{Tremonti04} used indirect estimations of gas masses, while in this work we are using direct measurements.  In this section, we
will show that such effect seems to be a consequence of AGN feedback.

\begin{figure*}[h]
\begin{center}
\includegraphics[scale=0.646]{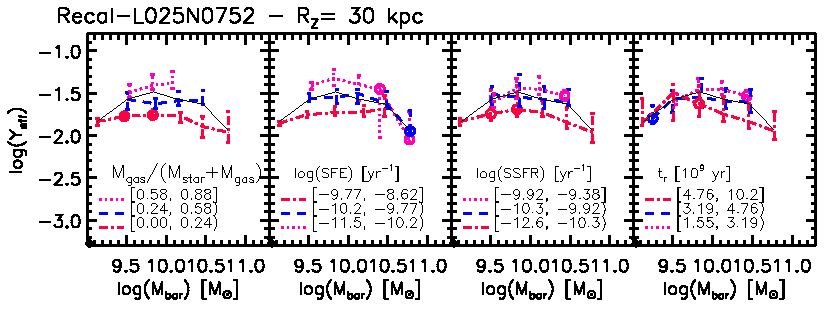}\\
\includegraphics[scale=0.646]{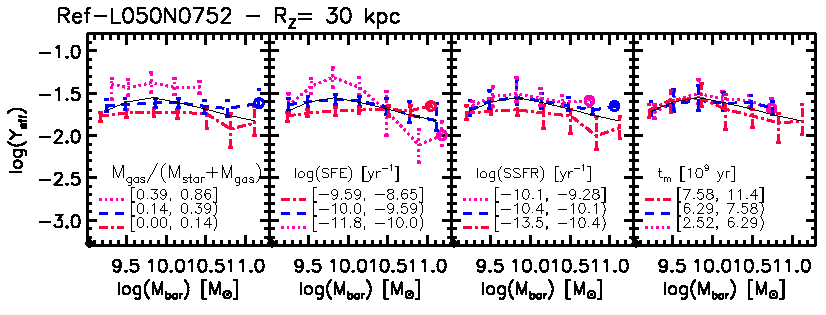}\\
\includegraphics[scale=0.646]{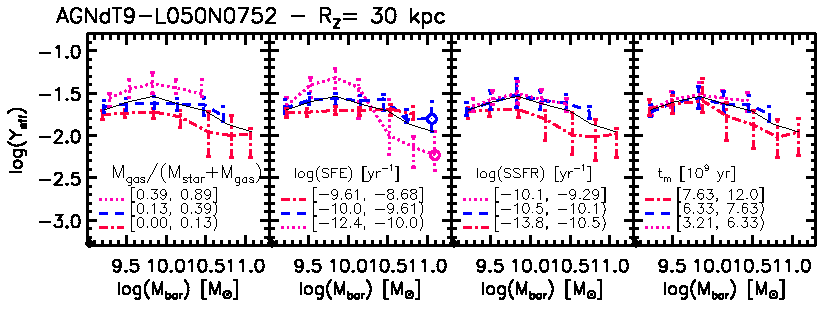}\\
\includegraphics[scale=0.646]{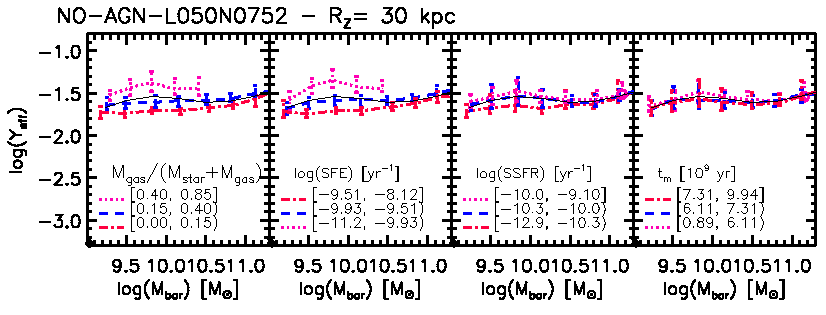}
\caption{ 
	Effective yields as a function of baryonic mass for different {\sc EAGLE} simulations: 
Recal-L025N0752, Ref-L050N0752, AGNdT9-L050N0752, NO-AGN-L050N0752 (from top to bottom,
as indicated in the figure).
The curves with error bars
indicate the $z=0$ median relations together with the 25th and 75th percentiles
corresponding to simulated galaxies binned in $f_{\rm gas}=M_{\rm gas}/(M_{\rm gas}+M_{\rm star})$,
SFE, SSFR and stellar mean age (from left to right, as indicated in the figure).
All considered mass bins contain $N_{\rm bin} \ge 5$ galaxies; less populated bins ($ 5 \le N_{\rm bin} < 10$) are marked with a circle.
As a reference, the median relation associated to the whole sample of galaxies is shown with
a black solid curve in each panel. Yields in this figure were derived by measuring metallicities inside
an aperture of $R_Z = 30$ kpc.  See Appendix \ref{Ap2}, for an analysis of aperture effects.
}
\label{eagle_comparison}
\end{center}
\end{figure*}

Fig.~\ref{eagle_comparison} show effective yields as a function of baryonic mass 
for different {\sc EAGLE} simulations:
Recal-L025N0752, Ref-L050N0752, AGNdT9-L050N0752 and NO-AGN-L050N0752 (from top to bottom,
as indicated in the figure).
Median relations at $z=0$ are shown binned in $\mu={\rm M}_{\rm gas}/({\rm M}_{\rm gas}+{\rm M}_{\rm star})$,
SFE, SSFR and stellar mean age (from left to right, as indicated in the figure).
All considered mass bins contain $N_{\rm bin} \ge 5$ galaxies; less populated bins ($ 5 \le N_{\rm bin} < 10$) are marked with a circle.
For simulation Recal-L025N0752, stellar mean age is weighted by luminosity in the $r$ band
using photometric data of stellar particles available in the public {\sc EAGLE} data base
\citep{mcalpine2015}.  As photometric data are not available for all `L050N0752' simulations,
we use the mass-weighted mean ages ($t_{\rm m}$) when analysing samples of galaxies extracted from
those simulations.

The top panels in Fig.~\ref{eagle_comparison} show results for the Recal-L025N0752 simulation.
 In each panel, the black solid line depicts the median relation associated to the whole
sample of galaxies. It is clear that,
for smaller galaxies, $y_{\rm eff}$ seems to correlate with 
M$_{\rm bar}$, on average. In addition, at a given log(${\rm M}_{\rm bar}) \la 10 \ {\rm M}_{\sun}$, 
systems with lower values of $y_{\rm eff}$ show lower $f_{\rm gas}$, higher SFEs and lower SSFRs;
but, there is a no clear dependence of $y_{\rm eff}$ on age at low masses.
On the other hand, for more massive galaxies, $y_{\rm eff}$ decreases with \Mbar\ on average.
In addition, at a given log(\Mbar) $\ga$ 10\Msun, lower values of $y_{\rm eff}$ are associated to old galaxies with
low gas fractions, SFEs and SSFRs; thus, passive systems whose SF process is quenched.

In order to investigate at what extent AGN feedback might be responsible for
the very low values of $y_{\rm eff}$ at high masses, we analyse simulations Ref-, AGNdT9- and NO-AGN-L050N0752.
As mentioned, the `Ref' model is the standard one, while the `NO-AGN' model suppresses all AGN feedback effects.  
On the other hand, the `AGNdT9' model predicts a stronger impact of AGN feedback on the chemical evolution of
galaxies, as discussed in detail in \citealt{derossi2017} (see their Sec. 3.3 and 7.1.2).
The three lower rows of panels in Fig.~\ref{eagle_comparison} compare the
\Mbar $ - y_{\rm eff}$ relations for the aforementioned simulations.
It is clear
that, at low masses, similar trends are obtained for the different AGN feedback prescriptions.
As expected, low mass systems seem not to be significantly affected by AGN feedback.  However, as mass
increases, AGN feedback effects seem to be stronger, leading to an anti-correlation between $y_{\rm eff}$ and
$M_{\rm bar}$ at the high-mass end.  Simulations run with the `Ref' and `AGN-dT9' model, which
include AGN feedback, predict such anti-correlation, while the `NO-AGN' model, in
which all BH processes were turned off, predicts a {\em correlation} between
$y_{\rm eff}$ and $M_{\rm bar}$ at high masses.  Moreover, the simulation `AGN-dT9',
which implements a higher temperature increment associated to AGN
heating ($\Delta T_{\rm AGN}=10^9 \ {\rm K}$), leads to a stronger decrease of $y_{\rm eff}$
at high masses. According to our results, systems more affected by AGN feedback
are old and exhibit very low $y_{\rm eff}$, $\mu$, SFE and SFR.  Thus, AGN feedback
seems to have quenched their SF and chemical evolution.  Only gas-rich mergers or 
significant gas accretion events might lead to further evolution of such galaxies.

Our findings for massive galaxies are consistent with previous results by \citet{derossi2017}.  According to these
authors, at a given mass, SFR tends to decreases as $\Delta T_{\rm AGN}$ increases because
the less frequent but more energetic feedback events corresponding to a higher $\Delta T_{\rm AGN}$
are more efficient at regulating SFR in massive galaxies.  In addition, at a given mass,
a higher $\Delta T_{\rm AGN}$ leads to lower metallicity values. \citet{derossi2017} conclude that
this behaviour seems to be driven by AGN feedback through three different channels:
the ejection of metal-enriched material, which generates a net metal depletion in simulated
galaxies; the heating of gas remaining in galaxies, which quenches the SF activity and, thus,
prevents further chemical enrichment from star formation; and, a minor role of net
metal dilution after star formation is suppressed by AGNs.


\section{On the relationship between \Mbar\ \lowercase{and}  \lowercase{y$_{\rm eff}$}}\label{MbarYieldSection}

\begin{figure*}[h]
\begin{center}
\includegraphics[scale=0.6]{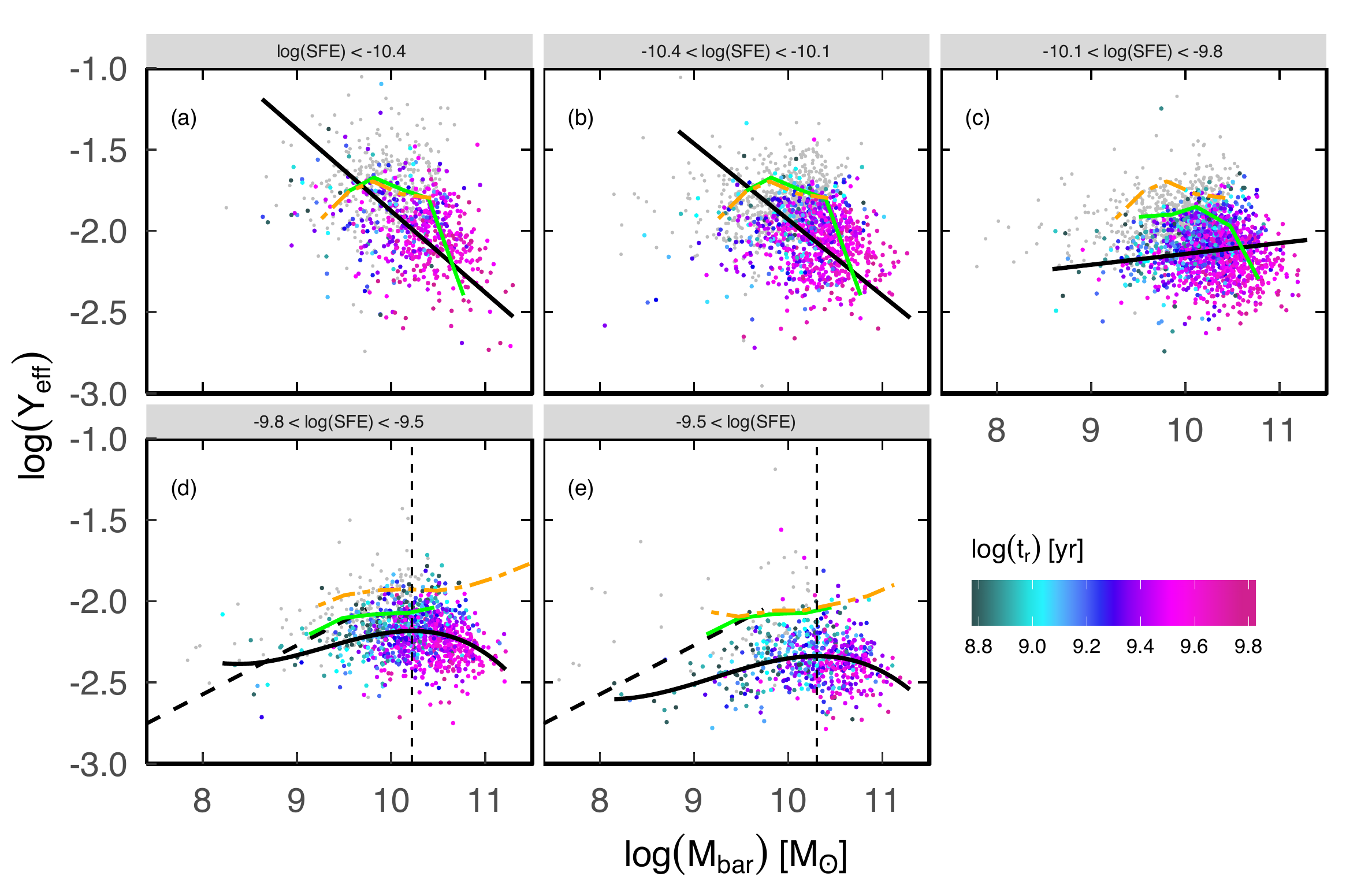}
\caption{Baryonic mass as a function of effective oxygen yield divided in panels of SFE. Every panel is color coded by age using a common color scale shown in the bottom right panel. The black solid lines show the linear and 3rd order polynomial fit to each panel, respectively, with its respective equations given in Eq. \ref{MbarYield_SFEAge}. The solid green lines correspond to the Recal-L025N0752 simulations of Fig. \ref{eagle_comparison} for the same variables. The dot dashed orange lines correspond to the No-AGN-L050N0752 models from the same figure. The dashed black line correspond to the relation of Ekta et al. for XMD galaxies (Eq. \ref{EktaEq}).  Vertical dashed lines show the point of inflection log(\Mbar) $\sim$ 10.22 for panel (d), and  log(\Mbar) $\sim$ 10.3 for panel (e). Only galaxies from the SDSS and ALFALFA surveys with measured age are shown (4266 galaxies).}
\label{MbarYieldPanelsSFE_Age}
\end{center}
\end{figure*}

With the aim of characterizing the driver of the opposite trends between low and high \Mbar\  galaxies, we analyze  in more detail the M$_{\rm bar}-$y$_{\rm eff}$ relation for different cuts in SFE, color coded by age, as shown in Fig \ref{MbarYieldPanelsSFE_Age}.  From top left to bottom right, all the black linear and polynomial fits from Fig. \ref{MbarYieldPanelsSFE_Age} are shown in Eq. \ref{MbarYield_SFEAge}, with its respective RMSE. The dashed lines in panels (d) and (e) show the fit of Ekta for XMD galaxies (Eq. \ref{EktaEq}),  while the vertical dashed lines in  panels (d) and (e) show the inflection point (stellar mass of the maximum value of the fitted function), in  which the data change from a correlation to an anti-correlation, log(\Mbar) $\sim$ 10.2 for panel (d), and  log(\Mbar) $\sim$ 10.3 for panel (e).\\

\begin{equation}\label{MbarYield_SFEAge}
       \begin{split}
& {\rm log(y_{\rm eff})}=  -0.5013 (\pm 0.0559) \times {\rm log(M_{bar}) } + 3.134 (\pm 0.5769) \\
& {\rm for} \;  {\rm   log(SFE) < -10.4,  \;  with }  \ RMSE = 0.25\\
& {\rm log(y_{\rm eff})}=  -0.4663 (\pm 0.0546) \times {\rm log(M_{bar}) } + 2.732 (\pm 0.5626) \\
& {\rm for} \;  {\rm -10.4 < log(SFE) < -10.1,  \;  with} \  \ RMSE = 0.24\\
& {\rm log(y_{\rm eff})}=  -0.06996 (\pm 0.13371) \times {\rm log(M_{bar}) } + 2.842 (\pm 1.373) \\
& {\rm for} \;  {\rm -10.1 < log(SFE) < -9.8,  \;  with } \  \ RMSE = 0.16\\
& {\rm log(y_{\rm eff})}= 47.4476 (\pm 17.1622) -16.3741 (\pm 5.2340) \times {\rm log(M_{bar}) } +  \\
& 1.7786 (\pm 0.5309) \times {\rm log(M_{bar})^2 } -0.0638 (\pm 0.0179) \times {\rm log(M_{bar})^3 } \\
& {\rm for} \;  {\rm -9.8 < log(SFE) < -9.5,  \;  with } \  \ RMSE = 0.13\\
& {\rm log(y_{\rm eff})}= 34.9975 (\pm 17.3501) -12.5641 (\pm 5.3152) \times {\rm log(M_{bar}) } +  \\
& 1.3836 (\pm 0.5414) \times {\rm log(M_{bar})^2 } -0.0501 (\pm 0.0183) \times {\rm log(M_{bar})^3 } \\
& {\rm for} \;  {\rm -9.5 < log(SFE),  \;  with }\  \ RMSE = 0.14\\
       \end{split}
\end{equation}

For comparison, we are including as well the EAGLE simulation model `Recal' in green lines, and `NO-AGN' in orange dot-dashed lines. The simulations shown in each panel correspond to the closest SFE from the models. Panels (a) and (b), show the simulation `Recal' with log(SFE) in the range  [-11.5 , -10.2],  panel (c) in the range  [-10.2 , -9.7], while panels (d) and (e) show the range  [-9.7 , -8.62]. As for the `NO-AGN' simulation, panels (a), (b), and (c) show log(SFE) in the range  [-11.2 , -9.93],  panel (d) shows the range [-9.93, -9.51], and  panel (e)  the range [-9.51, -8.12]. 

From Fig.  \ref{MbarYieldPanelsSFE_Age}, panels (a) and (b) show a clear anti-correlation between \Mbar\ and y$_{\rm eff}$. Furthermore, galaxies with such lower values of SFE are dominated mostly by old galaxies. Both panels have a median age of  log(t$_{\rm r}$) $\sim$ 9.5.  The EAGLE simulations show as well an anti-correlation in both, the `Recal' and `NO-AGN' models for the same panels. What we could conclude, is that for older galaxies the effective yield will decrease as the baryonic mass increases. 

For galaxies with $-10.1 <$ log(SFE) $< -9.8$ (panel c), there is no clear tendency, the median age for them is log(t$_{\rm r}$) $\sim$ 9.4, and corresponds to a rather transitional stage of galaxies. The next panels, (d) and (e), show intervals with a higher SFE, and show a relation for log(\Mbar)  $<$ 10, that anti-correlates for higher baryonic masses. Galaxies with log(\Mbar)  $>$ 10 are clearly older galaxies with median  log(t$_{\rm r}$) $\sim$ 9.4 and 9.3 for panels (d) and (e), respectively. Contrary, galaxies with log(\Mbar)  $<$ 10 are young galaxies with median  log(t$_{\rm r}$) $\sim$ 9.1 and 9.0 for panels (d) and (e), respectively. The simulation `Recal' agrees with a correlation for lower baryonic masses, however, it does not show any values for higher masses because of its limited associated simulated box (25 cMpc side lenght). On the other hand, the `NO-AGN' model shows that the correlation would continue for more massive galaxies if galaxies had not experience any AGN activity in their history.


From this figure, we can deduce that galaxies that are converting their gas into stars less efficiently (e.g., log(SFE) $<$ -10.4),  are older, and show a clear anti-correlation between their oxygen yield and baryonic mass. 
On the other hand, galaxies that are efficient at converting their mass into stars (e.g., log(SFE) $>$ -9.8) show a bi-modality that is separated by age, in which the yield increases with \Mbar \ for younger galaxies up to log(\Mbar) $<$10, and decreases for older galaxies for log(\Mbar) $>$10.


\begin{figure*}[h]
\begin{center}
\includegraphics[scale=0.6]{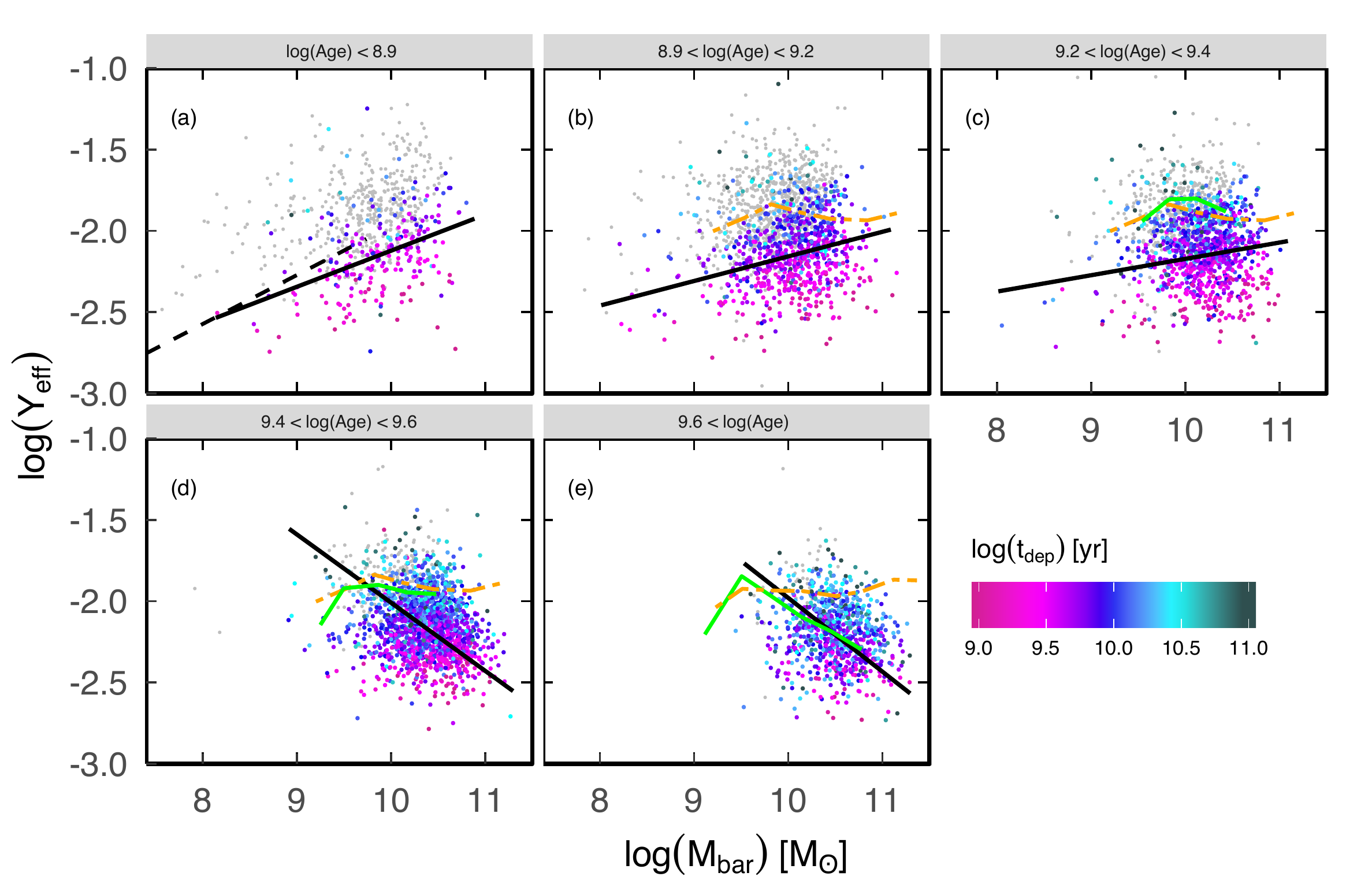}
\caption{Baryonic mass as a function of effective oxygen yield divided in panels of age. Every panel is color coded with log(t$_{\rm dep}$) using a common color scale shown in the bottom right panel. The black solid lines show the linear fit to each panel, with its respective equations given in Eq. \ref{MbarYield_AgeDepT}. The solid green lines correspond to the Recal-L025N0752 simulations of Fig. \ref{eagle_comparison} for the same variables. The dot dashed orange lines correspond to the No-AGN-L050N0752 models from the same figure. The dashed black line correspond to the relation of Ekta et al. for XMD galaxies (Eq. \ref{EktaEq}).}
\label{MbarYieldPanelsAge}
\end{center}
\end{figure*}

Since the stellar age is one of the key properties in our scaling relations, in Fig. \ref{MbarYieldPanelsAge} we show the M$_{\rm bar}$-y$_{\rm eff}$ relation for different cuts of age, color coded by their depletion time. All symbols are similar to the previous figure. As a comparison, we are showing as well the Recal  and NO-AGN simulations in green and orange lines, respectively. Panel (c) shows the `Recal' simulation for the age range of [9.19,  9.5] yr, panel (d) for [9.5, 9.67] yr, and panel (e) for [9.67,10], respectively. Unfortunately, the EAGLE simulations do not have a high enough resolution to appropriately describe younger, low mass galaxies, for which panels (a) and (b) do not show any simulations. On the other hand, the `NO-AGN' model for an age range of [8.9, 9.78] is shown in panels (b), (c) and (d), while panel (e) shows the age range [9.78, 9.86]. Similarly to the `Recal' simulations, the resolution is not good enough to describe younger low mass galaxies.  The fits to the observational data for the M$_{\rm bar}$-y$_{\rm eff}$ relation in bins of age (Fig. \ref{MbarYieldPanelsAge}) are listed below:

\begin{equation}\label{MbarYield_AgeDepT}
       \begin{split}
& {\rm log(y_{\rm eff})}=  0.2217 (\pm 0.0366) \times {\rm log(M_{bar}) } - 4.339 (\pm 0.363) \\
& {\rm for} \;  {\rm   log(t_r) < 8.9,  \;  with }\  \ RMSE = 0.23 \\
& {\rm log(y_{\rm eff})}=  0.1515 (\pm 0.0196) \times {\rm log(M_{bar}) } - 3.672 (\pm 0.197) \\
& {\rm for} \;  {\rm 8.9 < log(t_r) < 9.2,  \;  with}\  \ RMSE = 0.22\\
& {\rm log(y_{\rm eff})}=  0.09963 (\pm 0.03303) \times {\rm log(M_{bar}) } - 3.168 (\pm 0.337) \\
& {\rm for} \;  {\rm 9.2 < log(t_r) < 9.4\;  with}\  \ RMSE = 0.21\\
& {\rm log(y_{\rm eff})}=  - 0.4195 (\pm 0.0399) \times {\rm log(M_{bar}) } + 2.186 (\pm 0.414) \\
& {\rm for} \;  {\rm 9.4 < log(t_r) < 9.6\;  with}\  \ RMSE = 0.21\\
& {\rm log(y_{\rm eff})}=  - 0.4532 (\pm 0.0539) \times {\rm log(M_{bar}) } + 2.553 (\pm 0.567) \\
& {\rm for} \;  {\rm 9.6 < log(t_r)\;  with}\  \ RMSE = 0.21\\
       \end{split}
\end{equation}

In Fig. \ref{MbarYieldPanelsAge}, the stellar age provides a clear separation between the correlation and anti-correlations of the M$_{\rm bar}$-y$_{\rm eff}$ relation. For galaxies with a log(t$_{\rm r}$) $<$ 8.9, there is a clear correlation between those variables, the depletion times, i.e., the conversion of gas into stars, is fast, and there is a very good agreement with the Ekta et al. sample for XMD galaxies. The next panels (b), and (c), show how this correlation starts to flatten. For the same panels, the simulations `Recal' and `NO-AGN' show a similar tendency with a somehow higher yield.  For panels (d) and (e), we have an anti-correlation, in which the y$_{\rm eff}$ decreases as \Mbar \ increases. Furthermore, the depletion times increase with age, with a median log(t$_{\rm r}$) of 9.6, 9.8, 9.9, 9.97, and 10.1 for panels (a) to (e), respectively. The `Recal' simulation (green line) for the panel (e) agrees quite well with the observations. Again, the model for `NO-AGN' shows how the M$_{\rm bar}$-y$_{\rm eff}$ relation would continue to increase when no AGN is present in their history.

\section{Discussion}\label{Discussion}

The mass of a galaxy is critical for the fate of its evolution. Stars in more massive galaxies tend to have formed earlier and over shorter timescales; contrary to stars in low mass galaxies, which formed later and on longer timescales. Recently, the interest in this well established observational result has grown, and its implications have been more and more appreciated. This downsizing effect  \citep[e.g.,][]{Thomas05, Noeske07, Fontanot09} is still a matter of several studies. Indeed, a downsizing effect is invoked to explain the differences between opposite mass population of galaxies, however, its nature and physical explanation still remains unclear \citep[e.g.,][]{Neistein06}.

 Several authors have explored the properties of galaxies of opposite masses, for instance, \citet{Kauf03c}  show that low-redshift galaxies divide into two distinct families at a stellar mass of 3 $\times$ 10$^{10}$ M$_{\sun}$. Lower-mass galaxies have young stellar populations and low surface mass densities. A significant fraction of them have experienced recent starbursts, and the efficiency of the conversion of baryons into stars in low-mass galaxies increases in proportion to halo mass, perhaps as a result of supernova feedback processes. At stellar masses above 3 $\times$ 10$^{10}$ M$_{\sun}$, there is a rapidly increasing fraction of galaxies with old stellar populations, high surface mass densities and the high concentrations typical of bulges. They suggest that the star formation efficiency decreases in the highest--mass haloes, and that little star formation occurs in massive galaxies after they have assembled.

Furthermore, by analyzing the HI and H2 content in galaxies,  \citet{Saintonge11} found that for high mass galaxies, the molecular and atomic gas depletion timescales are comparable, but in low mass galaxies, molecular gas is being consumed much more quickly than atomic gas. Furthermore, scaling relations such as the M-SSFR \citep[e.g.,][]{Noeske07} clearly show that low mass galaxies are actively forming stars at a higher rate than massive galaxies per unit mass. 

 In is worth nothing that in comparison with previous works that divide galaxies in two distinct families based on their stellar mass, we find that the stellar age is a more clear divider in the properties of galaxies.

 Most of the scaling relations analyzed in this paper show smooth relations or anticorrelations, with RMSE values around 0.22 (M$_{\rm bar}$-E(B-V)$_{\rm gas}$, Fig. \ref{DustMbar} ), RMSE = 0.29 (t$_r$-SSFR, Fig. \ref{AgePlots}), and RMSE > 0.3 (e.g., \Mstar-SFE, Fig. \ref{SFERel}). On the other hand,  the M$_{\rm bar}$-y$_{\rm eff}$ relationships (Fig. \ref{MbarYieldPanelsAge}) shows a clear bi-modal behavior when galaxies are separated  by ($i$) stellar age, with RMSE values going from 0.21 to 0.23, and ($ii$) by SFE (Fig. \ref{MbarYieldPanelsSFE_Age}) , with RMSE values from 0.14 to 0.25.




Fig. \ref{MbarYieldPanelsAge} offers a clear picture of how galaxies are evolving. For one hand, young galaxies show faster depletion times, and a clear correlation between M$_{\rm bar}$ and y$_{\rm eff}$. This behavior has also been observed by several authors, for instance, \citet{Thuan16} analyzed as well the M$_{\rm bar}$-y$_{\rm eff}$ relation for BCD galaxies, finding a very similar result (see Fig. 13 of their paper), where there is a correlation for log(\Mbar) $<$ 10\Msun, which disappears at higher \Mbar, although suggesting a possible anti-correlation for higher masses that is not clear due to a low statistic. Similarly, other authors have analyzed this relation for low mass/metallicity galaxies, showing a very similar correlation  \citep[e.g.,][]{Garnett02, Ekta10}. \\

Among the possible explanations of why the effective yield increases with baryonic mass, it has been argued by 
\citet{Garnett02}, \citet{Tremonti04} and \citet{Silich01}, that the presence of galactic winds remove metals more efficiently from the shallower potential wells of low mass galaxies. \citet{SA14, SA15}  have argued that low metallicities in XMD galaxies are an indicator of infall of pristine gas, a process that would increase the effective yield in low-mass galaxies. Similarly, \citet{Ekta10} propose that the mechanism responsible for the metal deficiency of XMD galaxies, is a better mixing in the ISM, where low-metallicity gas from large galactocentric radii dilutes the central, metal-rich gas \citep[see also][]{Peeples09, Kewley06,Lee04}.  Furthermore, recently  \citet{Belfiore19} found that the inferred outflow loading factor decreases with stellar mass. It is likely that a combination of all of them to a different degree will explain the observed increase of the effective yield with baryonic mass we observe. We reiterate that for our sample, we recover a clean correlation  between M$_{\rm bar}$ and y$_{\rm eff}$ only when young galaxies are selected, as shown in Fig. \ref{MbarYieldPanelsAge}(a).


On the other hand, older galaxies show longer depletion times, and a clear anti-correlation between M$_{\rm bar}$ and y$_{\rm eff}$. EAGLE simulations indicate that black hole feedback is important for massive galaxies (\Mbar\ $>$ $10^{10}$\Msun), and a crucial effect to take into account to produce the mentioned anti-correlation.  Without black hole feedback in the history of these massive galaxies, today their effective yield would be way higher than the one observed, as shown in Fig \ref{MbarYieldPanelsAge} (d) \& (e). AGN feedback would have quenched the star formation and metallicity in old, massive galaxies, which is consistent with the observed low y$_{\rm eff}$, SFE, SSFR, and gas fractions. Again, the stellar age, and not just \Mbar\ is crucial, since young galaxies with \Mbar\ $>$ 10$^{10}$ still show a correlation between  M$_{\rm bar}$ and y$_{\rm eff}$. Another mechanism at play is the AGN gas heating, as mentioned in Sect \ref{sec_eagle_AGN}, by increasing $\Delta T_{\rm AGN}$ to 10$^9 \ {\rm K}$,  a stronger decrement in the y$_{\rm eff}$ is produced. On top of the AGN feedback effect, the depletion times for older galaxies tend to be longer compared to low-mass galaxies.

 The effect of an active AGNs in the history of current SF galaxies has been recently studied by \citet{Matthee19} using EAGLE simulations, finding that at masses M$_{\rm star}$ $>$ 10$^{10}$ M$_{\rm sun}$, the residuals in the M-SFR relation are anticorrelated with the relative efficiency of past Black Hole growth, which is a proxy of the accumulated AGN feedback energy. 

 Additionally, another possible effect to bear in mind was explored by \citet{Vincenzo16}, who find that high values of the effective yield may be indicative of an IMF favouring massive stars \citep{Kroupa01, Chabrier03}, and/or  a high mass cutoff of the IMF itself.

 It is important to keep in mind the various systematic effects that could bias our results. For example, the 3" diameter optical fiber used to take the SDSS spectra could bias our estimated metallicity by providing only information of the central part for large angular size galaxies. However, the EAGLE simulations show a very good agreement with our observed metallicity, for which this effect is probably not introducing a strong systematic bias. In this sense, \citet[][see their Appendix A]{derossi2017}  show that aperture effects does not affect the main trends found for metallicity scaling relations in EAGLE simulations. However, changes in the aperture can generate moderate variations of the slope and normalization of those relations,  for a more detailed discussion, refer to Appendix \ref{Ap2}.

 Additionally, \citet{Iglesias16} analyzed  aperture  corrections to a sample of SDSS galaxies. They find that 
for the redshift range 0.02 $<$ z $<$ 0.05, the average difference between fiber-based and integrated gas metallicities (estimated through the O3N2 index) would be of 0.034, 0.035, 0.037, 0.013, and 0.013 for the following stellar mass ranges in log(\Mstar / \Msun): 8.5$-$9.1, 9.1$-$9.7, 9.7$-$10.3, 10.3$-$10.9, 10.9$-$11.5, respectively. Even though the metallicity method is different than the one we are using in this paper, the largest metallicity over estimation would occur when  log(\Mstar / \Msun) $<$ 10.3.

Another important bias could be arise from the different spatial resolution between our optical and 
radio data, since the Arecibo L-band feed array has a bean size of 3.5 arcminutes, quite larger in 
comparison with the 3" fiber used by SDSS. However, our data again show a good agreement with the EAGLE 
simulations when using different apertures to measure SF gas associated to HII regions (R$_{\rm ap}$ = 30 kpc)  
and hydrogen mass (R$_{\rm ap}$= 70 kpc, Sec. 4.1).  
Furthermore, we have performed additional tests adopting the same aperture (30 kpc  and 70 kpc)  
when estimating the latter simulated quantities, obtaining similar general trends to those reported 
here for the M$_{\rm bar}$ -- y$_{\rm eff}$ relation, with only moderate changes in the slope and 
zero-point (see Appendix \ref{Ap2} ).  In this context, it is also worth emphasizing that 
De Rossi et al. (2017, see their figure 7) were able to reproduce the observed anti-correlation between 
y$_{\rm eff}$ and M$_{\rm bar}$ at high masses by measuring these properties within 30 kpc and 
considering only the total gas mass associated to the star-forming component of the ISM.   
 Finally, in Appendix \ref{Ap2}, we performed several tests, re-calculating properties for EAGLE
galaxies using different apertures.  Such results suggest that the very general trends reported 
here are robust against 
aperture effects: 1. There is a correlation between $M_{\rm bar}$ and $y_{\rm eff}$ at low masses. 
2. There is an anti-correlation between $M_{\rm bar}$ and $y_{\rm eff}$ at high masses. 3. At a fix mass, 
$y_{\rm eff}$ shows secondary dependencies on gas fraction, SFE, SSFR and age.


\begin{figure*}[h]
\begin{center}
\includegraphics[scale=0.3]{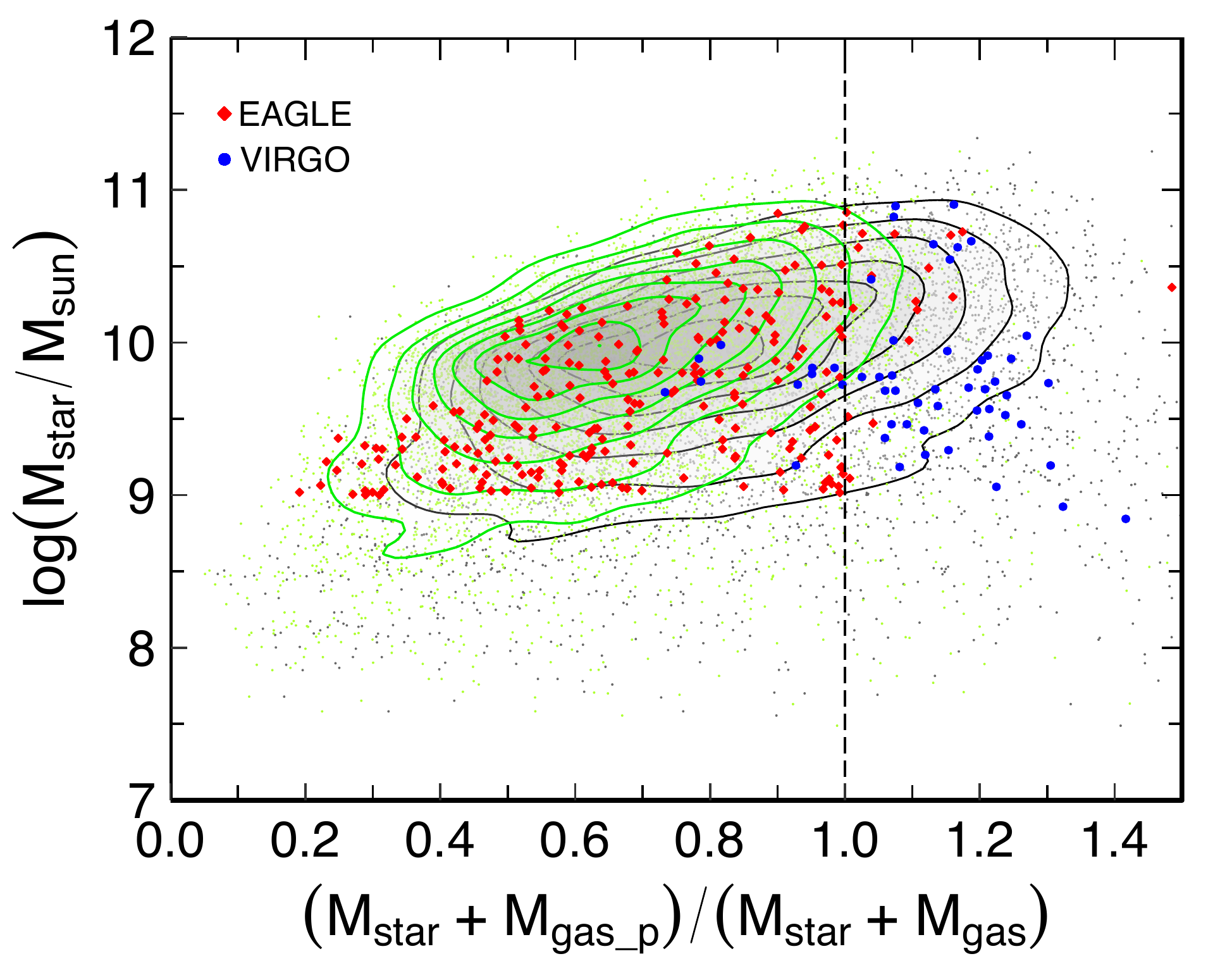}
\includegraphics[scale=0.3]{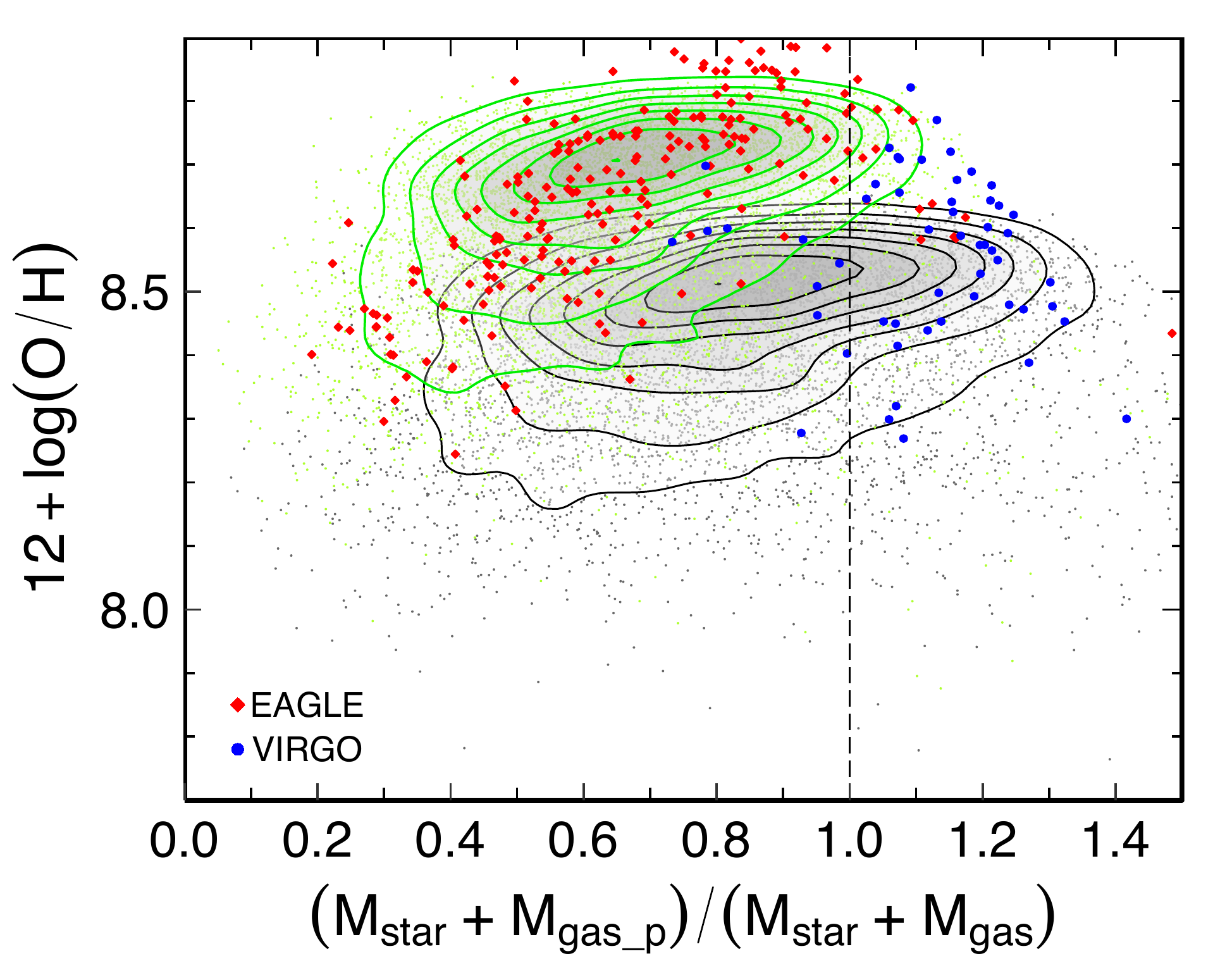}
\caption{Mass fraction of the protocloud estimated as in Eq. \ref{EqProtoCloud} vs. stellar mass (left), and gas metallicity (right). Green contours show the protocloud estimations using the measured metallicity, while black contours show the upper limit 12+log(O/H) - 0.2, assuming a typical gradient of $\sim$ 0.2 dex/R25 \citep[e.g.,][]{Pilyugin19}. Red diamonds and blue circles show data from the Recal EAGLE simulations and the Virgo cluster, respectively.}
\label{ProtoCloud}
\end{center}
\end{figure*}

 In order to quantify the extent of the circumgalactic medium (CGM) and the fiber effect of the SDSS spectra, we consider an approach to measure the fraction of the protocloud originally involved in the chemical evolution of the galaxy. By considering an inside-out scenario for galaxy formation \citep[e.g.,][]{Perez13}, we can do a rough estimate of the original protocloud assuming a closed-box model. First, we use Eq. \ref{EqYield} to estimate the gas fraction of the protocloud involved in the chemical evolution assuming a ${y}_{\rm true}$ $\sim$ 0.00268 \citep[][]{Pil04}, and the current gas metallicity provided by SDSS. Next, we use Eq \ref{GasFracEq} to solve for the gas mass of the protocloud (M$_{\rm gas\_p}$). We define the baryonic mass fraction of the protocloud involved in he chemical evolution as:

\begin{equation}\label{EqProtoCloud}
 {\rm \mu}_{\rm p} = ({\rm M_{gas}} + \rm{M_{star}}) \; / \; ({\rm M_{gas\_p}} + \rm{M_{star}})
\end{equation}

In Fig. \ref{ProtoCloud} we show $\mu_{\rm p}$ as a function of the stellar mass and gas metallicity. It is well known that galaxies show a gradient in gas metallicity from the center to the outskirts of galaxies. The typical value of the radial abundance gradient is $\sim$ 0.2 dex/R25 \citep[e.g.,][]{Pilyugin19}. Therefore, the 3" fiber used to obtain the SDSS spectra introduces a bias towards the center of the SDSS galaxies, thus, providing  systematically higher metallicities for some galaxies. In order to establish an upper and lower limit for this systematic effect, we estimate the mass fraction of the protocloud as indicated in Eq. \ref{EqProtoCloud} for two cases: ($i$) the observed metallicity (Fig. \ref{ProtoCloud}, green contours), and ($ii$) for  12+log(O/H) $-$ 0.2 (Fig. \ref{ProtoCloud}, black contours). In the same figure, red diamonds and blue circles correspond to galaxies from the EAGLE simulation Recal, and the Virgo cluster, respectively.  Note that in comparison with \citet{Iglesias16}, we are using a larger difference of $\sim$ 0.2 dex.

It is clear from  Fig. \ref{ProtoCloud} that for most galaxies the fraction of the protocloud is lower than the total baryonic mass. Furthermore, there is a linear trend, where $\mu_{\rm p}$ increases with stellar mass. This  approximation supports the idea that low mass galaxies have a gas rich CGM. On the other hand, galaxies on the right hand side of the dashed line in Fig. \ref{ProtoCloud}, show galaxies with a deficit of baryonic material. The most likely scenario is that galaxies located here have suffer from removal of gas  due to processes such as ram pressure stripping. Therefore, environmental effects must play a major role here. In fact, $\sim$82\% of galaxies from the Virgo cluster are systematically located here.

\section{Summary and Conclusions}\label{Conclusion}

In this paper we have combined Optical and Radio data from the SDSS and ALFALFA surveys to create a statistical significant sample. Moreover, we have used additional data from the Radio surveys GASS, COLD GASS, and a sample from the Virgo survey with Optical and Radio information. Through this paper we focused only on star forming galaxies selected with the BPT diagram. Also, we applied a completeness limit criteria to build a robust and reliable sample of galaxies. We analyzed scaling relations between several Optical (e.g., SFR, gas metallicity), and combined Optical/Radio (e.g., y$_{\rm eff}$, SFE, t$_{\rm dep}$, \Mbar) properties of galaxies. Additionally, we have compared our results with those of the EAGLE simulations.  A summary of our findings is listed here:

\begin{itemize}

\item We find that $\sim$98$\%$ of our SF sample has higher y$_{\rm eff}$ than the closed-box model prediction of log($y_0$)= $-2.57$ \citep{Pil04}.

\item Scaling relations with gas fraction show a smooth increment with SSFR, and an anti-correlation with SFE. This suggests that galaxies with high gas fractions, are producing a high amount of stars per unit of stellar mass, although are less efficient at transforming mass into stars in comparison with galaxies with low gas fractions. 

\item The SFE relationships with M$_{\rm star}$ and \Mbar-SFE are very different. While  the SFE increases linearly with M$_{\rm star}$, the SFE suggests a bimodality with  \Mbar\ (within the high dispersion), where galaxies with low \Mbar\ show a higher SFE that decreases as \Mbar\  increases, while galaxies with a high \Mbar\  show a high dispersion.

\item We use the color excess  E(B-V)$_{\rm gas}$ as indicative of extinction, finding a smooth increasing relation when log(\Mbar) $>$ 9.5. For lower values of \Mbar, this relation seems to flatten at E(B-V)$_{\rm gas}$ $\sim$ 0.15, suggesting lower amounts of dust for these galaxies. 

\item We analyzed the effective oxygen yields finding a bimodal distribution with \Mbar, suggesting a relation between them when log(\Mbar) $\lesssim$10, in good agreement with the XMD galaxies from \citep{Ekta10}. For higher values of \Mbar, our data suggest an anti-correlation.

    \item To gain a deeper understanding of our observational results, we use  a set of four different EAGLE simulations: Recal-L025N0752, Ref-L050N0752, AGNdT9-L050N0752, and NO-AGN-L050N0752. We compare directly the results of the Recal-L025N0752 simulation with our scaling relations, finding an excellent agreement. Indeed, we were able to even reproduce the bimodality observed in the \Mbar-y$_{\rm eff}$ relation.

\item To obtain a better interpretation of the SFE behavior for galaxies of different masses, we analyzed the \Mbar-y$_{\rm eff}$ relations in bins of SFE. We find that the SFE do not offer a clear bimodality of this relation. Indeed, for low SFE, galaxies are predominantly old, and there is an anti-correlation between \Mbar-y$_{\rm eff}$. However, galaxies with a high SFE (log(SFE) $> -9.8$), show a clear bimodality for the same relation, where there is an indication of a correlation for  log(\Mbar) $<$ 10, and an anti-correlation for higher masses. A comparison with the EAGLE simulations Recal shows a very good agreement. The model NO-AGN, in which AGN feedback is suppressed, is not able to reproduce the drop in y$_{\rm eff}$ observed for the higher SFE bins. Therefore, the inclusion of AGN feedback is very important to reproduce the observations.

\item We analyzed in more detail the \Mbar-y$_{\rm eff}$ relationship for a subsample of galaxies with direct measurements of stellar age. We find  that this relationship, with an apparent high dispersion, can be separated in clean correlations and anti-correlations when bins of stellar age are taken. For young galaxies, the  y$_{\rm eff}$ increases with \Mbar, furthermore, XMD galaxies from \citet{Ekta10} are in good agreement with this sub-sample. As the bins in stellar age increase, the \Mbar-y$_{\rm eff}$ relation starts to flatten until it becomes an anti-correaltion for log(t$_{\rm r}$) > 9.4 yr. The `Recal' simulation predicts the correlations and anti-correlation observed in this relation. Furthermore, the simulations predict that young, low mass systems are not significantly affected by AGN feedback. However, as age and baryonic mass increase, AGN feedback effects seem to be stronger, leading to an anti-correlation. Furthermore,  the `NO-AGN' model, predicts a correlation between the same variables.

 \item By using a rough estimation of the original protocloud of the galaxy, we find that low mass galaxies have a gas rich CGM. Furthermore, the CGM decreases as galaxies convert gas into stars, thus increasing their stellar mass. We find that the baryonic mass of the protocloud can be used to identify galaxies that have experienced intense loss of baryonic material.

  \item Among the possible reasons of why the effective yield changes, it is likely that several mechanisms are playing a role, and that their  degree of impact depend strongly on the age and baryonic mass the galaxies. 
     \begin{itemize}

    \item  For young galaxies with predominantly  low \Mbar,  it has been argued that the presence of galactic winds would remove metals more efficiently from the shallower potential  wells of low mass galaxies,  which would result in a higher  y$_{\rm eff}$. Furthermore, infall of pristine gas would increase the effective yield as well. A combination of both processes it is likely to produce a correlation between y$_{\rm eff}$ and \Mbar. Additionally, this tendency is stronger for young galaxies, which also show faster depletion times. \\

     \item For old galaxies, predominantly with log(\Mbar) $\gtrsim$ 10\Msun, the most important mechanism at play is likely to be a past AGN feedback effect, which would have quenched the SFR and chemical evolution of the galaxy. In this case, only gas-rich mergers or significant gas accretion events might lead to further evolution. The EAGLE simulation Recal is able to reproduce the observed anti-correlation between y$_{\rm eff}$ and \Mbar, also, the NO-AGN model is not able to reproduce this anti-correlation. Furthermore, the EAGLE simulation AGNdT9, which implements a higher temperature increment associated to AGN heating ($\Delta T_{\rm AGN}=10^9 \ {\rm K}$), leads to a stronger decrease of  y$_{\rm eff}$ at higher \Mbar. Therefore, different AGN heating temperatures could be affecting and decreasing  the y$_{\rm eff}$ in different degrees. 
\end{itemize}

     \item Placing our results in the downsizing scenario, young galaxies show predominantly low masses (although can extend up to \Mbar $\sim$ 10$^{10.5}$ \Msun), show faster depletion times, and do not seem to have been affected by an active AGN in their history. On the other hand, older galaxies are  predominantly massive (\Mbar $>$ 10$^{10}$\Msun\,), show on average longer depletion times, and AGN feedback must have played an important role in their history by quenching their star formation rate.

\end{itemize}

\section*{Acknowledgements}
We thank Shane O'Sullivan for insightful discussion. We also thank the anonymous referee
for their careful reading of this manuscript and helpful comments.
MEDR acknowledges support from PICT-2015-3125 of ANPCyT, PIP 112-201501-00447 of CONICET and UNLP G151 of UNLP (Argentina). AG acknowledges support by the INAF PRIN-SKA2017 program 1.05.01.88.04.ESKAPE-HI. TMH acknowledges the support from the Chinese Academy of Sciences (CAS) and the National Commission for Scientific and Technological Research of Chile (CONICYT) through a CAS-CONICYT Joint Postdoctoral Fellowship administered by the CAS South America Center for Astronomy (CASSACA) in Santiago, Chile.
IAZ acknowledges the support of the National Academy of Sciences of Ukraine by the grant 417Kt.
Funding for the SDSS and SDSS-II has been provided by the Alfred P. Sloan Foundation, the Participating Institutions, the National Science Foundation, the U.S. Department of Energy, the National Aeronautics and Space Administration, the Japanese Monbukagakusho, the Max Planck Society, and the Higher Education Funding Council for England. The SDSS Web Site is http://www.sdss.org/.
We acknowledge the Virgo Consortium
for making their simulation data available. The EAGLE
simulations were performed using the DiRAC-2 facility at Durham,
managed by the ICC, and the PRACE facility Curie based in France
at TGCC, CEA, Bruy\`{e}res-le-Ch\^{a}tel. 
  This work used the DiRAC@Durham facility managed by the Institute for
   Computational Cosmology on behalf of the STFC DiRAC HPC Facility
   (www.dirac.ac.uk). The equipment was funded by BEIS capital funding
   via STFC capital grants ST/P002293/1, ST/R002371/1 and ST/S002502/1,
   Durham University and STFC operations grant ST/R000832/1. DiRAC is
   part of the National e-Infrastructure.
We have benefited from the public available  ``R" statistical programing language (http://www.R-project.org/), and the TOPCAT analysis tool (http://www.starlink.ac.uk/topcat/).








\appendix

\section{Comparison of SFR\lowercase{s} and gas metallicities with other databases}\label{Ap1}

 With the aim of identify possible biases affecting our SFRs, in this section we perform a comparison between  the SFRs used in this paper by \citet{Brinchmann04}, the SFR estimated by \citet{Duarte17} using correction prescribed from IFU data, and direct integrated IFU MaNGA-DR14 SFRs estimated by us.  We find 59 MANGA-SDSS/ALFALFA counterparts, and 25 common counterparts in the 3 catalogs.

\begin{figure}
\begin{center}
\includegraphics[scale=0.21]{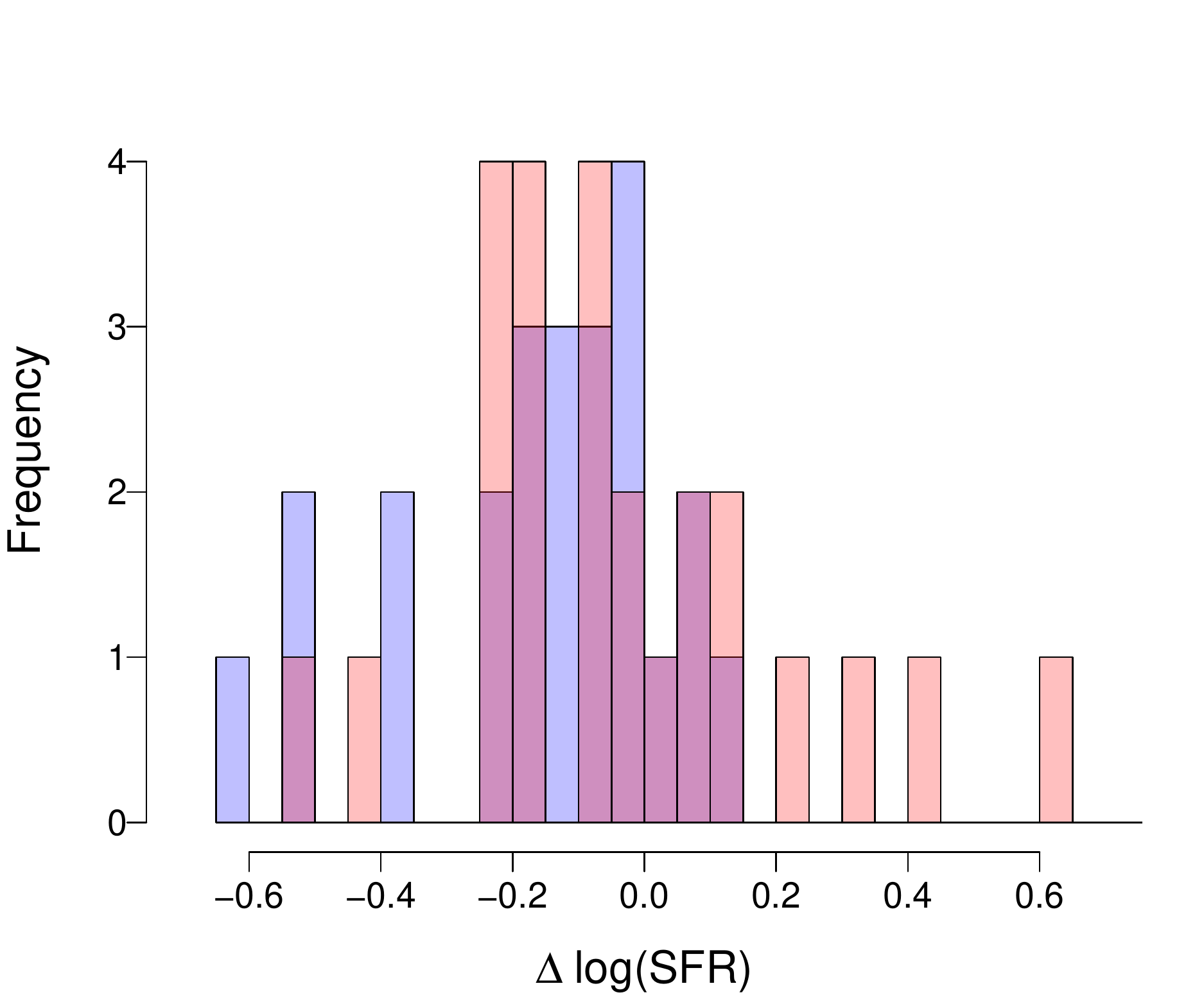}
\includegraphics[scale=0.21]{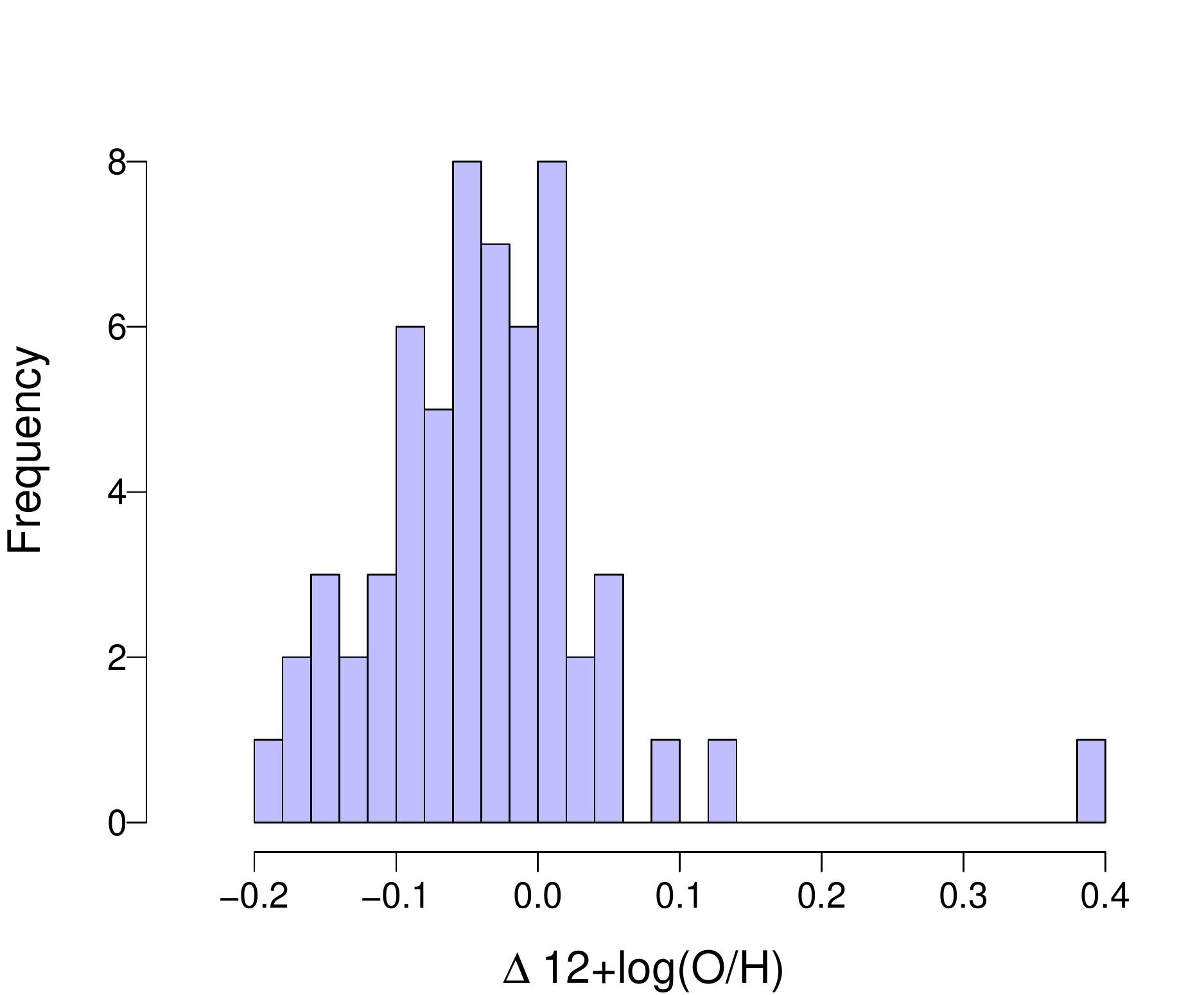}
\caption{Comparison between different SFRs and metallicities. Left: The histograms show the difference between MaNGA and  \citet{Brinchmann04} SFRs (blue), and  MaNGA and \citet{Duarte17} (red), for 25 common sources between the three catalogs. Right: The blue histogram shows the difference between integrated IFU MaNGA and 3 " fiber SDSS/ALFALFA gas metallicities estimated in this work for 59 common sources}
\label{SFRsComp}
\end{center}
\end{figure}

In Fig \ref{SFRsComp} left, we present a histogram of the log(SFR) difference between MaNGA and \citet{Brinchmann04} in blue, and MaNGA and \citet{Duarte17} in red for the 25 galaxies in common in the 3 catalogs. The mean difference between  the integrated log(SFR) of MaNGA and \citet{Brinchmann04}  is $\sim$0.026 dex, while for MaNGA and \citet{Duarte17}  is $\sim$0.03 dex. Furthermore, using the whole SDSS DR-7 sample, \citet{Ellison18} found 394 counterparts with MaNGA DR13 \citep{Albareti17}, finding that  the mean log(SFR) difference between \citet{Brinchmann04} and integrated MaNGA values is  $\sim$0.03 dex.

In Fig \ref{SFRsComp} right, we show a histogram with the gas metallicity difference between the 59 MaNGA and SDSS/ALFALFA counterparts. Metallicities were estimated consistently for the 2 data sets as indicated in Sect. 2.1 of this paper, using integrated IFU MaNGA data, and single 3" fiber for SDSS/ALFALFA data. We find that the  mean difference in 12+log(O/H) is $\sim$0.03 dex.

\section{\bf Simulated data using different apertures}\label{Ap2}

In this section, we present results from EAGLE simulations considering different apertures when
estimating galaxy properties.

\begin{figure*}
\begin{center}
\includegraphics[scale=0.646]{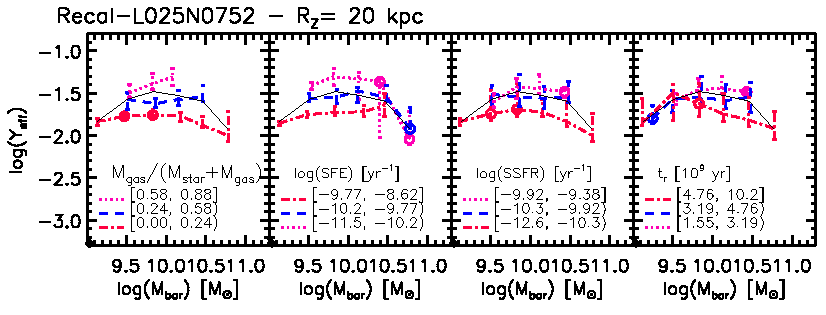}\\
\includegraphics[scale=0.646]{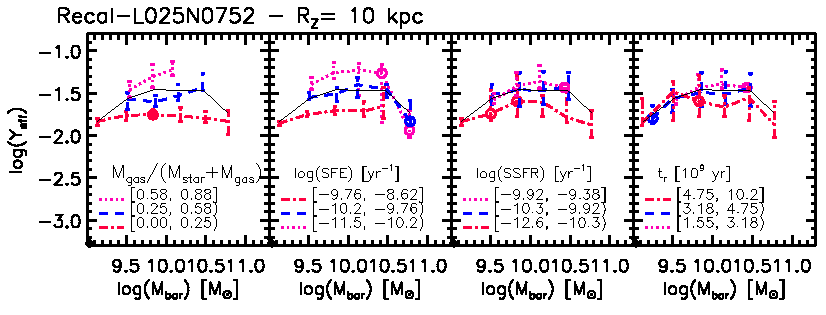}\\
\includegraphics[scale=0.646]{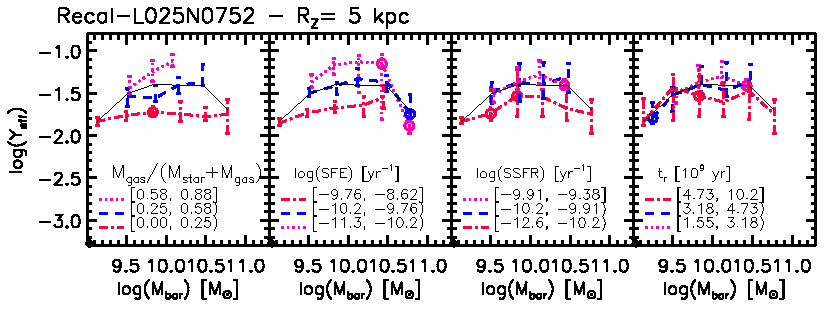}\\
\includegraphics[scale=0.646]{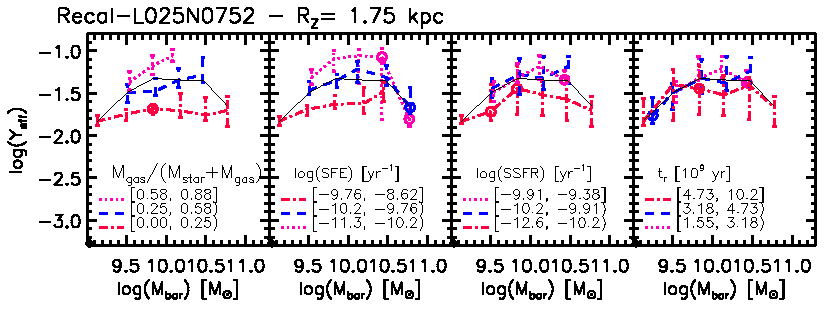}
\caption{ Effective yields as a function of baryonic mass for the EAGLE simulation Recal-L025N0752. 
	Metallicities here have been estimated using different fix apertures $R_Z$ (see title of each panel), 
	while all other properties were calculated using similar apertures to those used in Fig. 10 (top panel).
	Note that, for Fig. 10, $R_Z= 30$ kpc.}
\label{EAGLE_Recal_RZ}
\end{center}
\end{figure*}

In Fig. \ref{EAGLE_Recal_RZ}, all properties were calculated similarly to those shown in the 
top panel of Fig. \ref{eagle_comparison}, with the only exception of metallicities, which
were evaluated at different apertures $R_Z$, as shown in the different panels.
We consider a minimum $R_Z$ of $\approx 1.75$ kpc, which is close to the smallest apertures associated to 
metallicity estimates for the SDSS sample.
We see that a decrease of $R_Z$ generates a slight variation in the shape of the 
$M_{\rm bar}-y_{\rm eff}$ relation and a moderate increase of effective yields ($\lesssim 0.5$ dex),
with the most significant variations in the case of gas-rich galaxies.
However, it is clear, that the main trends of the $M_{\rm bar}-y_{\rm eff}$ relation are preserved
when varying $R_Z$: there is a correlation at the low-mass end and an aticorrelation at the high-mass
end. Also, at a fix mass, the secondary dependences of $y_{\rm eff}$ on gas fraction, SFE, SSFR and
$t_r$ are still present at small $R_Z$.

\begin{figure*}
\begin{center}
\includegraphics[scale=0.646]{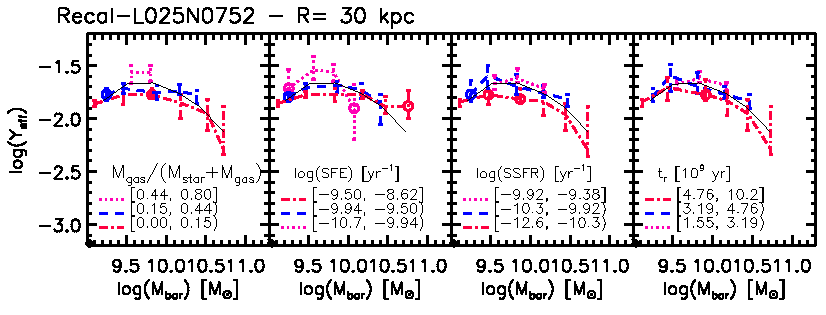}\\
\includegraphics[scale=0.646]{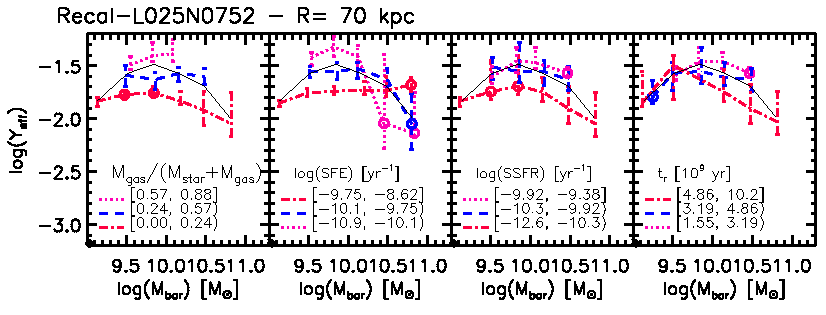}
	\caption{ Effective yields as a function of baryonic mass for the EAGLE simulation Recal-L025N0752. 
	In contrast with Fig. 10 (top panel), in the top panel here, all properties have been estimated using a fix aperture of 30 kpc.  In the bottom panel, all properties were estimated using a fix aperture of 70 kpc, with the only exception of the age ($t_r$). The luminosity-weighted age $t_r$ was estimated using an aperture of 
	30 kpc as the EAGLE database does not provide luminosities at larger apertures.}
\label{EAGLE_Recal30kpc}
\end{center}
\end{figure*}

In Fig. \ref{EAGLE_Recal30kpc}, we show results from EAGLE simulations when evaluating all galaxy properties
at the same fix aperture of 30 kpc (top panel) and 70 kpc (bottom panel).
When comparing with Fig. \ref{eagle_comparison} (top panels), we see that the main trends of the relationships 
are preserved, although variations in the slope and normalization are present. The most significant
variations are obtained when measuring all properties at $\lesssim 30$ kpc because of the decrease of
the total gas mass with respect to our {\em default} value for this quantity, which was measured
at $\lesssim 70$ kpc.  On the other hand, not very significant changes with respect
to Fig. \ref{eagle_comparison} are obtained when evaluating all quantities at $\lesssim 70$ kpc.
This is a consequence of the fact that the simulated stellar component and star-forming gas-phase (used
to estimate metallicities) tend to be located in the central regions of galaxies ($\lesssim 30$ kpc).

We conclude that our main findings are robust against aperture effects.

%


\bsp	
\label{lastpage}
\end{document}